\documentclass[a4paper,accepted=2021-07-10,10pt,twocolumn,]{quantumarticle}
\pdfoutput=1

\usepackage{amsmath}
\usepackage{amssymb}
\usepackage{bbm}
\usepackage{framed}
\usepackage{epstopdf}
\usepackage{color}
\usepackage[colorlinks = true, urlcolor = blue]{hyperref}
\usepackage{epstopdf}
\usepackage{dsfont}
\usepackage{amsthm}
\usepackage{wrapfig}
\usepackage{relsize}
\usepackage{bm}
\usepackage[latin1]{inputenc}
\usepackage{enumitem}
\usepackage[numbers]{natbib}
\usepackage{graphicx}
\usepackage{physics}
\usepackage{xcolor}
\usepackage{hyperref}
\newtheorem{thm}{Theorem}[section]
\newtheorem{lemma}{Lemma}[section]

\begin{document}

	\title{Partially Coherent Direct Sum Channels}

	\author{Stefano Chessa}
	\orcid{0000-0003-2771-8330}
	\email{stefano.chessa@sns.it}
	\author{Vittorio Giovannetti}
	\orcid{0000-0002-7636-9002}
	\affiliation{NEST, Scuola Normale Superiore and Istituto Nanoscienze-CNR, I-56126 Pisa, Italy}

	\date{2020-07-10}

\begin{abstract}
{We introduce \textit{Partially Coherent Direct Sum} (PCDS) quantum channels, as a generalization of the already known \textit{Direct Sum}  quantum channels.  We derive necessary and sufficient conditions to identify the
subset of those maps which are degradable, and  provide a simplified expression for their quantum capacities. Interestingly, the special structure of
 PCDS allows us to extend the computation of the  quantum capacity formula also for quantum channels which are 
 explicitly not degradable (nor antidegradable).   
We show instances of applications of the results to dephasing channels, amplitude damping channels, and combinations of the two.}
\end{abstract}

\maketitle

\section{Introduction} \label{sec.Intro} 

Since the seminal work of Shannon \cite{SHANNON} the performance analysis of means of communication has been rephrased into the search of the maximum information transmission rates achievable by noisy channels. In the last few decades the possibility of exploiting quantum-mechanical effects to perform communication has been brought to the light. The approach to the evaluation of channel capacities  was then directed also towards noisy quantum channels, opening the road to the field of quantum communication \cite{HOLEVOBOOK, WATROUSBOOK, WILDEBOOK, NC}. 

The features of quantum information can bring advantages w.r.t. classical settings, e.g., among others, quantum key distribution \cite{QKD1}, superdense coding \cite{SUPERDENSE}, quantum teleportation \cite{TELEPORT}, superactivation of the quantum capacity of a quantum channel \cite{SUPERACT},  superadditivity of the classical capacity of a quantum channel \cite{SUPERADD_CL}. The qualitatively different phenomena that typically are involved when dealing with quantum systems on the other hand might constitute challenging obstacles, see as an example the often intractable regularizations needed for the definitions of channel capacities \cite{HOLEGIOV,SURVEY}. Even for few uses of the channel considered, in absence of useful properties or further symmetries, maximizations over Hilbert spaces (to find capacities) can reveal themselves computationally hard, especially in higher dimensions. This makes the study, in terms of information capacities, of a wide realm of channels unattractive and unexplored, despite quantum information in higher dimensions being a well established field of research. For this family of systems and channels potential advantages are showed either from the quantum computation (see e.g. \cite{COMP,COMP1,COMP2,COMP3}) and quantum communication (see e.g. \cite{COMM,COMM1,COMM2,COMM3}) perspectives, and are now also increasingly experimentally accessible \cite{EXP,EXP1,EXP2,EXP3,EXP4,EXP5,EXP6,EXP7,EXP8}. All this considered, methods to overcome these kind of obstacles are still researched and this paper aims to contribute to this corpus of literature. Specifically, we present compact expressions for the quantum capacity and entanglement assisted quantum capacity of a new class of channels that we called \textit{Partially Coherent Direct Sum} (PCDS) channels, a generalization of the direct sum (DS) channels described in \cite{FUKUDA}. This formalism appears in a variety of contexts \cite{DS1, DS2, DS3, DS4, DS5, DS6, DS7}, among which recently the ``gluing'' procedure in \cite{DS8} derived as a generalization of the construction in \cite{DS9}. DS channels were initially introduced in \cite{FUKUDA} in the context of the additivity conjecture for the classical capacity, later proven wrong \cite{SUPERADD_CL}. There, the direct-sum structure was proposed in order to simplify the expressions of those functionals, e.g. output entropy and Holevo information, whose additivity was conjectured. Among these results also an expression for the coherent information for DS channels was found. In the subsequent literature the direct-sum structure has been mainly exploited to show superadditivity features of quantum channels. 

 We draw attention to PCDS because, other than encompassing the already known DS channels, they provide a framework for the efficient description of wide classes of physical noise models and their capacities. An instance is given by damping processes in multi-level systems, denominated multi-level amplitude damping (MAD) channels \cite{QUTRIT_ADC}. In addition to that, PCDS appear to be interesting because their capacity is in principle exactly computable with reduced complexity also for high dimensional systems. In this sense, the knowledge already acquired about low dimensional quantum channels can be exploited to compose new PCDS channels. At the same time the introduction of the PCDS can push the study of all accessible zoology of low dimensional channels. In addition to that, through the techniques here developed, in some cases we are able to evaluate exactly the quantum capacity even if the channel can be proven not to be degradable~\cite{DEVSHOR}. Finally we also see that PCDS channels, despite the similar construction, have higher capacities w.r.t. DS channels. This enhancement is
associated with the direct sum structure of the Hilbert space and the introduction of coherences.

The manuscript is organized as follows: in Sec.~\ref{Sec:MODEL} we introduce the model for the channels we consider; in Sec~\ref{sec:deg} we analyze complementary channels and degradability properties; in Sec.~\ref{SEC:COMP} we study the quantum capacity and entanglement assisted quantum capacity; in Sec~\ref{Sec:application} we apply results to instances of quantum channels that include dephasing channels, amplitude damping channels and combinations of the two. Conclusions and
perspectives are presented in Sec.~\ref{conclusione} while technical material is presented in the Appendix.

\section{The model} \label{Sec:MODEL} 

Let us start fixing some notation: 
given ${\cal H}_{\text{X}}$ and ${\cal H}_{\text{Y}}$ two Hilbert spaces associated with two
(possibly unrelated) quantum systems X and Y, we shall use the symbol
\begin{eqnarray} {\cal L}_{{\text{X}} \rightarrow {\text{Y}}}:= \{ \hat{\Theta}_{\text{YX}}: {\cal H}_{\text{X}} \longrightarrow {\cal H}_{\text{Y}}\}\;, \end{eqnarray} 
 to represent the set of linear operators $\hat{\Theta}_{\text{YX}}$
mapping the input vectors of X into the output vectors of Y.
The symbol  $\mathfrak{S}_{{\text{X}}}= \mathfrak{S}({\cal H}_{{\text{X}}})$ will describe the special  subset of $\cal{L}_{{\text{X}} \rightarrow {\text{X}}}$ formed by the density operators $\hat{\rho}_{\text{XX}}$ of the system X. 
We also define 
\begin{eqnarray} {\cal M}_{{\text{X}} \rightarrow {\text{Y}}}:= \{{\Phi}_{\text{YX}}: \cal{L}_{{\text{X}} \rightarrow {\text{X}}}
 \longrightarrow \cal{L}_{{\text{Y}} \rightarrow {\text{Y}}}\}\;, \end{eqnarray} 
 to be  the set of super-operators $\Phi_{\text{YX}}$ which
transform operators $\hat{\Theta}_{\text{XX}}\in \cal{L}_{{\text{X}} \rightarrow {\text{X}}}$ into elements  of $\cal{L}_{{\text{Y}} \rightarrow {\text{Y}}}$. We'll indicate with  ${\cal M}^{\text{(cpt)}}_{{\text{X}} \rightarrow {\text{Y}}}$ 
  the special subset  formed by the quantum channels of  ${\cal M}_{{\text{X}} \rightarrow {\text{Y}}}$, i.e. by the super-operators ${\Phi}_{\text{YX}}$  which are Completely Positive and Trace preserving (CPT) \cite{CPTP}. 
 Finally, for X$\neq$Y  we shall use the special symbol 
 \begin{eqnarray} {\cal M}^{\text{(off)}}_{\text{X}\rightarrow \text{Y}}:= 
\{{\Phi}^{\text{(off)}}_{\text{YX}}: \cal{L}_{{\text{X}} \rightarrow {\text{Y}}}
 \longrightarrow \cal{L}_{{\text{X}} \rightarrow {\text{Y}}}\}\;, \end{eqnarray} 
 to describe linear mappings ${\Phi}^{\text{(off)}}_{\text{YX}}$  which connect operators  $\cal{L}_{{\text{X}} \rightarrow {\text{Y}}}$
 into themselves.

Consider next C, a quantum system described by a Hilbert space ${\cal H}_{\text{C}}$  admitting the following direct sum decomposition 
\begin{eqnarray}
{\cal H}_{\text{C}}= {\cal H}_{\text{A}} \oplus {\cal H}_{\text{B}}\;,  \label{dec}
\end{eqnarray} 
with ${\cal H}_{\text{A}}$ and ${\cal H}_{\text{B}}$ two nontrivial  subspaces of dimensions
$d_{\text{A}}$, $d_{\text{B}}=d_{\text{C}}-d_{\text{A}}$. We'll associate with them the projectors ${\hat{P}}_{\text{AA}}$ and ${\hat{P}}_{\text{BB}}$
that fulfill the 
orthonormalization conditions 
\begin{eqnarray} 
\hspace{-20pt}{\hat{P}}_{\text{AA}} {\hat{P}}_{\text{BB}}={\hat{P}}_{\text{BB}}{\hat{P}}_{\text{AA}}=0 \;, \quad {\hat{P}}_{\text{AA}}+{\hat{P}}_{\text{BB}} = \hat{I}_{\text{CC}} \;, \label{ORTHO} 
\end{eqnarray} 
 $\hat{I}_{\text{CC}}$ being the identity operator on ${\cal H}_{\text{C}}$.
Accordingly, any operator 
${\hat{\Theta}}_{\text{CC}} \in {\cal{L}}_{{\text{C}} \rightarrow {\text{C}}}$ mapping  the space  of C into itself  can then be written  
 as a sum of diagonal and off-diagonal block terms, i.e. 
\begin{eqnarray} \label{D1} 
{\hat{\Theta}}_{\text{CC}}  =\bigoplus_{\text{X,Y}=  \text{A,B}}  {\hat{\Theta}}_{\text{YX}} \equiv 
\left[  \begin{array}{c|c}
{\hat{\Theta}}_{\text{AA}} & {\hat{\Theta}}_{\text{AB}}  \\ \hline 
{\hat{\Theta}}_{\text{BA}} & {\hat{\Theta}}_{\text{BB}} \end{array} \right],
\end{eqnarray} 
with ${\text{X,Y}}={\text{A,B}}$ and ${\hat{\Theta}}_{\text{XY}}$ is an element of ${\cal{L}}_{{\text{Y}} \rightarrow {\text{X}}}$ defined by the identity
\begin{eqnarray}\label{eq:op projections}
{\hat{\Theta}}_{\text{XY}}\equiv \hat{P}_{\text{XX}} {\hat{\Theta}}_{\text{CC}} \hat{P}_{\text{YY}}\;.
\end{eqnarray}

Let now $\Phi_{\text{CC}} \in {\cal M}^{\text{(cpt)}}_{{\text{C}} \rightarrow {\text{C}}}$ be a  CPT channel mapping C into itself. 
 We say that it is a \textit{Partially Coherent Direct Sum} (PCDS) map if it preserves the block structure in Eq.~(\ref{D1}). Or equivalently if we can identify super-operators 
$\Phi_{\text{AA}}\in {\cal M}_{{\text{A}} \rightarrow {\text{A}}}$, $\Phi_{\text{BB}}\in {\cal M}_{{\text{B}} \rightarrow {\text{B}}}$, $\Phi^{\text{(off)}}_{\text{AB}}\in {\cal M}^{\text{(off)}}_{{\text{B}} \rightarrow {\text{A}}}$, and $\Phi_{\text{BA}}^{\text{(off)}}\in {\cal M}^{\text{(off)}}_{{\text{A}} \rightarrow {\text{B}}}$ 
such that 
\begin{eqnarray} \label{D2} 
\Phi_{\text{CC}}\left[  \begin{array}{c|c}
{\hat{\Theta}}_{\text{AA}} &  {\hat{\Theta}}_{\text{AB}}  \\ \hline 
 {\hat{\Theta}}_{\text{BA}} &  {\hat{\Theta}}_{\text{BB}} \end{array} \right]= \left[  \begin{array}{c|c}
 \Phi_{\text{AA}}[{\hat{\Theta}}_{\text{AA}}]  &  \Phi^{\text{(off)}}_{\text{AB}}[{\hat{\Theta}}_{\text{AB}}] \\ \hline 
\Phi^{\text{(off)}}_{\text{BA}}[{\hat{\Theta}}_{\text{BA}}] & \Phi_{\text{BB}}[{\hat{\Theta}}_{\text{BB}} ]\end{array} \right]\;,\nonumber \\
\end{eqnarray} 
for all input $\hat{\Theta}_{\text{CC}}\in {\cal{L}}_{{\text{C}} \rightarrow {\text{C}}}$. In brief, simplifying the notation
\begin{eqnarray} \label{eq:channel decomposition}
\Phi_{\text{CC}} =  \Phi_{
\text{AA}} +  \Phi_{
\text{BB}} +    \Phi_{
\text{AB}}^{\text{(off)}} + \Phi_{
\text{BA}}^{\text{(off)}}\;,
 \label{BRIEF} 
\end{eqnarray}
where each channel $\Phi_{\text{XX}}$ implicitly assumes the projection on the suitable subspace as in Eq.~(\ref{eq:op projections}).
 Quantum channels that can be cast in form of Eq.~(\ref{BRIEF}) arise whenever the quantum system 
C is affected by a (possibly noisy) evolution that preserves the relative populations associated with 
the 
subsystems ${\cal H}_{\text{A}}$ and ${\cal H}_{\text{B}}$, but 
(possibly) deteriorates the  quantum coherence among them. 
 In Appendix \ref{APPANEC} it is shown a necessary and sufficient condition for this to happen. This condition
 is that, given $\{ {\hat{M}}_{\text{CC}}^{(j)}\}_j$ a Kraus set \cite{KRAUS} for 
 $\Phi_{\text{CC}}$,  its 
 elements must only involve diagonal terms when cast into the block form as in Eq.~(\ref{D1}), i.e.
 \begin{thm} 
A quantum channel $\Phi_{\text{CC}}$  described by a Kraus set $\{ {\hat{M}}_{\text{CC}}^{(j)}\}_j$
admits the PCDS structure~(\ref{BRIEF})   if and only if  
 \begin{eqnarray} \label{KRAUS} 
 {\hat{M}}_{\text{CC}}^{(j)} =  {\hat{M}}_{\text{AA}}^{(j)} +  {\hat{M}}_{\text{BB}}^{(j)}\;,  
 \end{eqnarray}
 or equivalently  that ${\hat{M}}_{\text{AB}}^{(j)} = {\hat{M}}_{\text{BA}}^{(j)} =0$, for all $j$. 
		\label{thm:necandsuff}
\end{thm}
The explicit proof of this result is given in Appendix~\ref{APPANEC}. There we also show that the maps on the right-hand-side of Eq.~(\ref{BRIEF}) can 
  be  expressed in terms of the operators 
 ${\hat{M}}_{\text{AA}}^{(j)}$ and ${\hat{M}}_{\text{BB}}^{(j)}$
  of  Eq.~(\ref{KRAUS}) as
\begin{eqnarray} 
\Phi_{\text{XX}} [\cdots]  &=&  \sum_j {\hat{M}}_{\text{XX}}^{(j)}  \cdots {\hat{M}}_{\text{XX}}^{(j)\dag}\;, \quad \
 \label{OFFDIAG} \end{eqnarray} 
for all X$=$A,B 
and 
\begin{eqnarray} 
\Phi^{\text{(off)}}_{\text{XY}} [\cdots]  &=&  \sum_j {\hat{M}}_{\text{XX}}^{(j)}  \cdots {\hat{M}}_{\text{YY}}^{(j)\dag}\;,\label{OFFDIAG1} 
\end{eqnarray} 
for all X$\neq$Y$=$A,B. Notice that in particular Eq.~(\ref{OFFDIAG}) implies that 
 the diagonal terms 
define proper  CPT channels on A and B  respectively, i.e. $\Phi_{\text{AA}} \in {\cal M}^{\text{(cpt)}}_{{\text{A}} \rightarrow {\text{A}}}$ and $\Phi_{\text{BB}} \in {\cal M}^{\text{(cpt)}}_{{\text{B}} \rightarrow {\text{B}}}$, with Kraus sets provided by 
$\{ {\hat{M}}_{\text{AA}}^{(j)}\}_j$ and $\{ {\hat{M}}_{\text{BB}}^{(j)}\}_j$.

One can easily check that given $\Phi'_{\text{CC}},\Phi''_{\text{CC}} \in {\cal M}^{\text{(cpt)}}_{{\text{C}} \rightarrow {\text{C}}}$ 
fulfilling the constraint of Eq.~(\ref{BRIEF}), then the same holds true for both the channel  $p \Phi'_{\text{CC}} + (1-p) \Phi''_{\text{CC}}$ with  $p\in [0,1]$
and for the channel  $\Phi'_{\text{CC}} \circ \Phi''_{\text{CC}}$,  with $``\circ"$ representing super-operator composition. 
The first property implies that the set of PCDS quantum evolutions is closed under convex combination. The second property instead,
together with the observation that the identity channel $\text{Id}_{\text{CC}}$ is also trivially PCDS, tells us that the set forms a semi-group under channel concatenation. 
Observe also that a special instance of PCDS transformations is provided  by the 
purely dephasing channels  $\Delta^{(\kappa)}_{\text{CC}}$, which induce the mapping 
\begin{eqnarray} \label{D2def} 
\Delta^{(\kappa)}_{\text{CC}}\left[  \begin{array}{c|c}
{\hat{\Theta}}_{\text{AA}} &  {\hat{\Theta}}_{\text{AB}}  \\ \hline 
 {\hat{\Theta}}_{\text{BA}} &  {\hat{\Theta}}_{\text{BB}} \end{array} \right]= \left[  \begin{array}{c|c}
{\hat{\Theta}}_{\text{AA}}  &  \kappa {\hat{\Theta}}_{\text{AB}} \\ \hline 
\kappa^* {\hat{\Theta}}_{\text{BA}} &{\hat{\Theta}}_{\text{BB}}\end{array} \right]\;,\nonumber \\
\end{eqnarray} 
with $\kappa$ being a complex parameter of norm $|\kappa|\leq 1$. 
The semi-group property mentioned above also allows us to state the following. Starting from any PCDS channel $\Phi_{\text{CC}}$, described as in Eq.~(\ref{BRIEF}) for some 
proper choice of the maps $\Phi_{\text{AA}}$, $\Phi_{\text{BB}}$, $\Phi^{\text{(off)}}_{\text{AB}}$, and $\Phi^{\text{(off)}}_{\text{BA}}$,  we can 
construct an entire family of new  PCDS elements 
 \begin{eqnarray} \label{defKappa} 
 \Phi_{\text{CC}}^{(\kappa)}\equiv  \Delta^{(\kappa)}_{\text{CC}} \circ \Phi_{\text{CC}} = 
 \Phi_{\text{CC}} \circ   \Delta^{(\kappa)}_{\text{CC}} \;, 
 \end{eqnarray}   whose off-diagonal components  are rescaled versions of 
$\Phi^{\text{(off)}}_{\text{AB}}$, and $\Phi^{\text{(off)}}_{\text{BA}}$, i.e. 
\begin{eqnarray} 
\Phi^{(\kappa)}_{\text{CC}} =   \Phi_{\text{AA}} +\Phi_{\text{BB}} +  \kappa   \Phi^{\text{(off)}}_{\text{AB}} + \kappa^*  \Phi^{\text{(off)}}_{\text{BA}}\;.
 \label{BRIEF1}  
\end{eqnarray}
(Here the commutativity property exhibited in Eq.~(\ref{defKappa}) follows from the linearity of the super-operators $\Phi^{\text{(off)}}_{\text{AB}}$
and $\Phi^{\text{(off)}}_{\text{BA}}$).
In particular by setting $\kappa=0$, Eq.~(\ref{BRIEF1}) describes   the 
\textit{Direct Sum} (DS) channels discussed in 
Ref.~\cite{FUKUDA} which  
 completely suppress coherence among ${\cal H}_{\text{A}}$ and ${\cal H}_{\text{B}}$.
 This special condition is  met whenever the Kraus elements in Eq.~(\ref{KRAUS})  of a PCDS map are given  by 
 operators that   have support exclusively either on ${\cal H}_{\text{A}}$ or on ${\cal H}_{\text{B}}$. We can summarize this constraint in terms of the following simple relation 
  \begin{eqnarray} \label{KRAUSk0} 
  {\hat{M}}_{\text{AA}}^{(j)} \neq 0  \Longrightarrow  M_{\text{BB}}^{(j)}=0\;,  \qquad \forall j\;. 
 \end{eqnarray} 
 It is worth stressing that the properties discussed above, as well as the results we are going to present in the following sections,
 admit a simple generalization in case of multi-block decompositions of the map PCDS -- see Appendix~\ref{GENAPP}. 
\section{Complementary channels and degradability conditions for PCDS maps} \label{sec:deg}

 We remind that, via the Stinespring dilation theorem \cite{STINE}, 
 given $\Phi_{\text{XX}}\in {\cal M}^{\text{(cpt)}}_{{\text{X}} \rightarrow {\text{X}}}$ a generic CPT transformation  on an arbitrary  system X, its complementary channel can be identified  with a
  CPT map  $\tilde{\Phi}_{\text{EX}}\in {\cal M}^{\text{(cpt)}}_{{\text{X}} \rightarrow {\text{E}}}$ 
 coupling X with the (sufficiently large) auxiliary quantum system E that plays the role of the system environment.
Let $\{ M_\text{XX}^{(j)}\}_j$ be a Kraus set for $\Phi_{\text{XX}}$ and $\{ |j_{\text{E}}\rangle\}_j$ be a fixed set of orthonormal vectors on the Hilbert space ${\cal H}_{\text{E}}$ of E.  Then the 
 action of  $\tilde{\Phi}_{\text{EX}}$ on a generic operator ${\hat{\Theta}}_{\text{XX}}\in{\cal L}_{{\text{X}} \rightarrow {\text{X}}}$  can be expressed as 
 \begin{eqnarray} \label{COMPPHI} 
\tilde{\Phi}_{\text{EX}}[{\hat{\Theta}}_{\text{XX}} ] = \sum_{j,j'} |j_{\text{E}}\rangle\!\langle j'_{\text{E}}| \; \mbox{Tr} \big[ M_{XX}^{(j')\dag} 
M_{XX}^{(j)} {\hat{\Theta}}_{\text{XX}} \big] \;,\nonumber \\
\end{eqnarray} 
(notice that due to the arbitrariness of the choice of $\{ |j_{\text{E}}\rangle\}_j$, $\tilde{\Phi}_{\text{EX}}$ can always be 
redefined up to a unitary rotation on E).
We also remind that the map $\Phi_{\text{XX}}$ is said to be degradable~\cite{DEVSHOR} if
we can identify a degrading CPT quantum channel $\Lambda_{\text{EX}}\in
{\cal M}^{\text{(cpt)}}_{{\text{X}} \rightarrow {\text{E}}}$ which allows us to reconstruct the action of  $\tilde{\Phi}_{\text{EX}}$
by acting on the corresponding output of ${\Phi}_{\text{XX}}$, i.e.  
\begin{eqnarray}\label{DEGCOND} 
\tilde{\Phi}_{\text{EX}} = \Lambda_{\text{EX}} \circ \Phi_{\text{XX}}\;. 
\end{eqnarray} 
Similarly we say that  $\Phi_{\text{XX}}$ is antidegradable \cite{ANTIDEGRADABLE} if exists 
 $\Lambda_{\text{XE}}\in
{\cal M}^{\text{(cpt)}}_{{\text{E}} \rightarrow {\text{X}}}$ such that 
\begin{eqnarray}
\Phi_{\text{XX}} = \Lambda_{\text{XE}} \circ \tilde{\Phi}_{\text{EX}}\;. 
\end{eqnarray}

In the case of DS channel ${\Phi}_{\text{CC}}$, using Eq.~(\ref{KRAUSk0}) and the orthogonality 
between ${\hat{M}}_{\text{AA}}^{(j)}$ and ${\hat{M}}_{\text{BB}}^{(j)}$, from Eq.~(\ref{COMPPHI}) 
one can then easily verify that for all
input operators ${\hat{\Theta}}_{\text{CC}}$ the following identity holds
\begin{eqnarray} 
\tilde{\Phi}_{\text{EC}}[{\hat{\Theta}}_{\text{CC}}] &=& \tilde{\Phi}_{\text{EA}}[{\hat{\Theta}}_{\text{AA}}] + 
\tilde{\Phi}_{\text{EB}}[{\hat{\Theta}}_{\text{BB}}] \;. \label{ERE12} 
\end{eqnarray}
 Here $\tilde{\Phi}_{\text{EA}}$ and $\tilde{\Phi}_{\text{EB}}$ are, respectively,  the complementary channels associated with the  diagonal components $\Phi_{\text{AA}}$ and $\Phi_{\text{BB}}$ entering in the decomposition of Eq.~(\ref{BRIEF}),
 while ${\hat{\Theta}}_{\text{AA}}$ and ${\hat{\Theta}}_{\text{BB}}$  are the diagonal terms of Eq.~(\ref{D1}).
 We refer the reader to  Appendix~\ref{APPA} for a physical insight on this identity. 
As we'll see next, Eq.~(\ref{ERE12}) holds also for PCDS. Notice though that, while for generic PCDS channels  $\Phi_{\text{CC}}$ the operators  $\tilde{\Phi}_{\text{EA}}[{\hat{\Theta}}_{\text{AA}}]$ 
 and $\tilde{\Phi}_{\text{EB}}[{\hat{\Theta}}_{\text{BB}}]$  may have nontrivial commutation relations,
 in the special case of the DS channels $\Phi^{(0)}_{\text{CC}}$~\cite{FUKUDA} they have always zero overlap, i.e.
 \begin{eqnarray}
  \tilde{\Phi}^{(0)}_{\text{EA}}[{\hat{\Theta}}_{\text{AA}}] 
\tilde{\Phi}^{(0)}_{\text{EB}}[{\hat{\Theta}}_{\text{BB}}] =  \tilde{\Phi}^{(0)}_{\text{EB}}[{\hat{\Theta}}_{\text{BB}}] 
\tilde{\Phi}^{(0)}_{\text{EA}}[{\hat{\Theta}}_{\text{AA}}] =0\;.\nonumber \\
 \end{eqnarray}
  In this scenario this implies that the sum appearing in Eq.~(\ref{ERE12}) is indeed a direct sum.  
  
We can now prove a necessary and sufficient condition for the degradability of a  generic PCDS channel  $\Phi_{\text{CC}}$. This condition establishes that such property only depends upon the diagonal blocks entering in the decomposition of Eq.~(\ref{BRIEF}): 

\begin{thm}
A PCDS  quantum channel $\Phi_{\text{CC}}$  is degradable if and only if all of its diagonal block terms  
$\Phi_{\text{AA}}$, $\Phi_{\text{BB}}$  are degradable too.
		\label{thm:optimalDeltaM}
\end{thm}
{\it Proof:} 
First of all let us show  that the degradability of $\Phi_{\text{AA}}$ and $\Phi_{\text{BB}}$ 
 implies the degradability of $\Phi_{\text{CC}}$.
Indeed  for $\text{X}=\text{A,B}$, let $\Lambda_{\text{XE}}$ be the CPT degrading map from  X  to E, which 
allows us to express $\tilde{\Phi}_{\text{XX}}$ in terms of $\Phi_{\text{XX}}$ as in Eq.~(\ref{DEGCOND}).
Consider then the super-operator $\Lambda_{\text{EC}}$ 
from  C  to E defined as \begin{eqnarray} 
\Lambda_{\text{EC}}[{\hat{\Theta}}_{\text{CC}}] \equiv  \label{DEFCON} 
 \Lambda_{\text{EA}}[{\hat{\Theta}}_{\text{AA}}]  +\Lambda_{\text{EB}}[ {\hat{\Theta}}_{\text{BB}}] \;,
\end{eqnarray} 
which is CPT thanks to the fact that both $\Lambda_{\text{EA}}$ and $\Lambda_{\text{EB}}$ fulfill the same constraint by hypothesis -- see Appendix~\ref{APPCPTCOND} for details. Furthermore for all ${\hat{\Theta}}_{\text{CC}}$  we have 
 \begin{eqnarray} 
&&\Lambda_{\text{EC}} \circ  \Phi_{\text{CC}} [{\hat{\Theta}}_{\text{CC}}] = \Lambda_{\text{EC}}  \left[  \begin{array}{c|c}
 \Phi_{\text{AA}}[{\hat{\Theta}}_{\text{AA}} ] & \Phi^{\text{(off)}}_{\text{AB}}[{\hat{\Theta}}_{\text{AB}}] \\ \hline 
 \Phi^{\text{(off)}}_{\text{BA}}[{\hat{\Theta}}_{\text{BA}}]  & \Phi_{\text{BB}}[{\hat{\Theta}}_{\text{BB}}]  \end{array} \right]   \nonumber \\
&&\qquad \qquad = \Lambda_{\text{EA}} \circ \Phi_{\text{AA}}[{\hat{\Theta}}_{\text{AA}} ] +  \Lambda_{\text{EB}} \circ
  \Phi_{\text{BB}}[{\hat{\Theta}}_{\text{BB}}  \big] 
  \nonumber \\
  &&\qquad \qquad =  \tilde{\Phi}_{\text{EA}}[{\hat{\Theta}}_{\text{AA}}] +   \tilde{\Phi}_{\text{EB}}[{\hat{\Theta}}_{\text{BB}}]= \tilde{\Phi}_{\text{EC}}[{\hat{\Theta}}_{\text{BB}} ] \;, \nonumber 
 \end{eqnarray} 
 that proves that $\Phi_{\text{CC}}$ is degradable with degrading channel as in Eq.~(\ref{DEFCON}). 

Let's now show next that if $\Phi_{\text{CC}}$ is degradable then also $\Phi_{\text{AA}}$ and $\Phi_{\text{BB}}$ must be degradable. 
For this purpose, given $\Lambda_{\text{EC}}$  the CPT transformation from  C  to E which allows us to reconstruct 
$\tilde{\Phi}_{\text{EC}}$ from $\Phi_{\text{CC}}$, from Eqs.~(\ref{BRIEF})  and (\ref{ERE12}) we get 
\begin{equation}
\tilde{\Phi}_{\text{EA}}[{\hat{\Theta}}_{\text{AA}}] + \tilde{\Phi}_{\text{EB}}[{\hat{\Theta}}_{\text{BB}}]= 
 \sum_{\text{X,Y}= \text{A,B}}
(\Lambda_{\text{EC}}\circ \Phi_{\text{YX}})[{\hat{\Theta}}_{\text{YX}}] \;,
\end{equation} 
which must hold true for all ${\hat{\Theta}}_{\text{YX}} \in {\cal L}_{{\text{X}} \rightarrow {\text{Y}}}$. 
In the particular case ${\hat{\Theta}}_{\text{BB}} = {\hat{\Theta}}_{\text{AB}} = {\hat{\Theta}}_{\text{BA}} =0$, this implies 
that for all ${\hat{\Theta}}_{\text{AA}}  \in {\cal L}_{{\text{A}} \rightarrow {\text{A}}}$ we have 
\begin{equation}
\tilde{\Phi}_{\text{EA}}[{\hat{\Theta}}_{\text{AA}}]=(\Lambda_{\text{EC}}\circ \Phi_{\text{AA}})[{\hat{\Theta}}_{\text{AA}}] 
= (\Lambda_{\text{EA}}\circ \Phi_{\text{AA}})[{\hat{\Theta}}_{\text{AA}}]\;,
\end{equation} 
where in the last identity we introduced 
\begin{eqnarray} \label{ddf} 
 \Lambda_{\text{EA}} [\cdots ] \equiv \Lambda_{\text{EC}}[ \hat{P}_{\text{AA}} \cdots \hat{P}_{\text{AA}}] \;,
\end{eqnarray} 
by exploiting the fact that $\Phi_{\text{AA}}$ maps operators of  A into A, i.e. that $ \hat{P}_{\text{BB}}\Phi_{\text{AA}}[{\hat{\Theta}}_{\text{AA}}]= \Phi_{\text{AA}}[{\hat{\Theta}}_{\text{AA}}] \hat{P}_{\text{BB}} =0$.
Since the map in Eq.~(\ref{ddf}) is CPT -- see Appendix~\ref{APPCPTCOND}, we can finally conclude that $\Phi_{\text{AA}}$ is degradable.
The degradability of $\Phi_{\text{BB}}$ can be proved in the same way.
$\square$

\section{Computing the quantum capacity of PCDS channels}  \label{SEC:COMP} 

As firstly shown in \cite{QCAP1,QCAP2,QCAP3}, the quantum capacity $Q(\Phi_{\text{XX}})$ of a channel $\Phi_{\text{XX}}$ is expressed as: 
\begin{equation} \label{DEFQUA} 
Q(\Phi_{\text{XX}})=\lim\limits_{n\rightarrow\infty} \frac{1}{n} \max_{\hat{\rho}^{(n)}_{\text{XX}}  \in \mathfrak{S}({\cal H}_\text{X}^{\otimes n})} I_{coh}(\Phi_{\text{XX}}^{\otimes n}; \hat{\rho}^{(n)}_{\text{XX}}), 
\end{equation}
where $I_{coh}(\Phi_{\text{XX}}; \hat{\rho}_{\text{XX}})$ is the \textit{coherent information} and is defined as 
\begin{equation} \label{COHERENT} 
I_{coh}(\Phi_{\text{XX}}; \hat{\rho}_{\text{XX}}) \equiv S( \Phi_{\text{XX}}(\hat{\rho}_{\text{XX}}))- S( \tilde{\Phi}
 _{\text{EX}}(\hat{\rho}_{\text{XX}}))\;,
\end{equation}
being $S(\hat{\rho}_{\text{XX}})\equiv- \text{Tr}_{\text{X}}\left[\hat{\rho}_{\text{XX}}\log_2 \hat{\rho}_{\text{XX}}\right]$ the von Neumann entropy and $\tilde{\Phi}
 _{\text{EX}}$ the complementary channel of $\Phi_{\text{XX}}$ as defined in Eq.~(\ref{COMPPHI}). As already mentioned in the introduction, the challenging aspect of the computation of the quantum capacity is given by the regularization over the number $n$ of channel uses. This since the behavior for many uses doesn't scale linearly w.r.t. the single shot formula, due to the well known property of non additivity of quantum channels. The issue can be bypassed when the channel is degradable (see Sec.~\ref{simply}) for which the single letter formula is sufficient \cite{DEVSHOR}, or antidegradable (the complementary channel is degradable) for which, due to no-cloning argument, we have $Q(\Phi_{\text{XX}})=0$.
Since we'll make use of this feature, it is finally worth noticing that from the 
invariance of the von Neumann entropy under unitary transformations it follows that  the capacity formula
 reported above does not depend on the specific form of the complementary channel. Indeed as already mentioned, the complementary channel can be chosen freely up to a unitary rotation  acting on the environment E -- see more about this in App.~\ref{APPA}.\\

Moving now towards DS and PCDS channels, in Ref.~\cite{FUKUDA} it was shown that the quantum capacity of DS channels is given by the maximum of the quantum capacity of their diagonal contributions. Expressed in our notation
\begin{eqnarray}\label{DSCAP} 
Q( \Phi^{(0)}_{\text{CC}}) =  \max \{ Q( \Phi_{\text{AA}}),Q( \Phi_{\text{BB}})\} \;,
\end{eqnarray} 
with $\Phi_{\text{AA}}$ and $ \Phi_{\text{BB}}$ being the diagonal block terms. 
The presence of non-zero off-diagonal contributions in Eq.~(\ref{BRIEF}) is clearly bound to challenge the above result.
To begin with, invoking the channel \textit{data-processing inequality} (DPI)~\cite{HOLEVOBOOK, WATROUSBOOK, WILDEBOOK,KEYL,WILDE1,NC} from Eq.~(\ref{defKappa}), it follows that the right-hand-side of Eq.~(\ref{DSCAP}) 
is an explicit lower bound for the quantum capacity of  an arbitrary PCDS channel 
$\Phi_{\text{CC}}$ having the same diagonal block terms of  $\Phi^{(0)}_{\text{CC}}$ , i.e. 
\begin{eqnarray}  \label{trivialbound} 
Q( \Phi_{\text{CC}}) \geq  Q( \Phi^{(0)}_{\text{CC}})  =\max \{ Q( \Phi_{\text{AA}}),Q( \Phi_{\text{BB}})\} \;.\nonumber \\
\end{eqnarray} 
This fact alone paves the way to higher communication performances.  The easiest way to see this is by  comparing the 
case of the identity map ${\text{Id}}_{\text{CC}}$, which has capacity 
\begin{eqnarray} Q( {\text{Id}}_{\text{CC}}) =\log_2 d_{\text{C}} =\log_2 (d_{\text{A}} +d_{\text{B}})  \;, \label{QK1} 
\end{eqnarray}  
with the case of the completely dephasing channel $\Delta^{(\kappa=0)}_{\text{CC}}$ of Eq.~(\ref{D2def}). $\Delta^{(\kappa=0)}_{\text{CC}}$  shares the same diagonal terms of ${\text{Id}}_{\text{CC}}$ (i.e. $\Phi_{\text{AA}}= {\text{Id}}_{\text{AA}}$
and  $\Phi_{\text{BB}}= {\text{Id}}_{\text{BB}}$) but, according to Eq.~(\ref{DSCAP}),
 has instead  quantum capacity equal to  
 \begin{eqnarray} \label{QK0} 
 Q( \Delta^{(\kappa=0)}_{\text{CC}}) =\max\{ \log_2 d_{\text{A}},\log_2 d_{\text{B}}\}\;.
 \end{eqnarray} 
Exploiting the results of the previous section we are going to set this observation on a broader context, computing the explicit
value of the quantum capacity of  large class of PCDS channels. Interestingly enough this will allow us to determine the quantum capacity of channels which are not degradable. 

\subsection{The quantum capacity of degradable PCDS channels} \label{simply} 

Consider the case of a PCDS channel  $\Phi_{\text{CC}}$ which is degradable. According to~\cite{DEVSHOR} we can
hence express it in terms of the following single-letter expression 
\begin{eqnarray}\label{QCPA} 
Q( \Phi_{\text{CC}}) = \max_{\hat{\rho}_{\text{CC}}\in \mathfrak{S}_{{\text{C}}}} 
I_{coh}(\Phi_{\text{CC}}; \hat{\rho}_{\text{CC}})  \;,
\end{eqnarray} 
with $I_{coh}(\Phi_{\text{CC}}; \hat{\rho}_{\text{CC}})$ the single-use coherent information functional introduced in Eq.~(\ref{COHERENT}). 
Observe next that from Eq.~(\ref{D2}) and the monotonicity of $S$ under block diagonalization, it follows that 
\begin{eqnarray} 
&& \hspace{-11pt}S(\Phi_{\text{CC}}(\hat{\rho}_{\text{CC}}))
 = S\left(   \left[  \begin{array}{c|c} p  \Phi_{\text{AA}}[\hat{\tau}_{\text{AA}}]  &  \Phi^{\text{(off)}}_{\text{AB}}[\hat{\rho}_{\text{AB}}]\nonumber \\ \hline 
\Phi^{\text{(off)}}_{\text{BA}}[\hat{\rho}_{\text{BA}}] &  (1-p) \Phi_{\text{BB}}[\hat{\tau}_{\text{BB}}] \end{array} \right]\right) \nonumber \\
 &&
 \quad  \label{ineq1} \leq S\left(   \left[  \begin{array}{c|c}
p  \Phi_{\text{AA}}[\hat{\tau}_{\text{AA}}]  & 0\\ \hline 
0& (1-p) \Phi_{\text{BB}}[\hat{\tau}_{\text{BB}}] \end{array} \right]\right)\nonumber \\
&&\quad  = S \left(  
p  \Phi_{\text{AA}}[\hat{\tau}_{\text{AA}}]\right)   +  S\left( (1-p) \Phi_{\text{BB}}[\hat{\tau}_{\text{BB}}] \right)  \nonumber \\
&&\quad  \nonumber = p S\left(  
  \Phi_{\text{AA}}[\hat{\tau}_{\text{AA}}]\right)   + (1-p) S\left( \Phi_{\text{BB}}[\hat{\tau}_{\text{BB}}] \right)  + H_2(p)\;.\nonumber \\
\end{eqnarray} 
Here we fixed $p\equiv \mbox{Tr} [ \hat{\rho}_{\text{AA}}]$, 
we introduced the density matrices of A and B defined as $\hat{\tau}_{\text{AA}}= \hat{\rho}_{\text{AA}}/p$ and $\hat{\tau}_{\text{BB}}= \hat{\rho}_{\text{BB}}/(1-p)$, and 
called $H_2(p)\equiv - p \log_2 p - (1-p) \log_2 (1-p)$ the binary entropy function. 
Considering then that Eq.~(\ref{ineq1}) can be saturated by focusing on density matrices 
$\hat{\rho}_{\text{CC}}$ with zero off-diagonal blocks (i.e. $\hat{\rho}_{\text{AB}}=\hat{\rho}_{\text{BA}}=0$),
and using the fact that according to Eq.~(\ref{ERE12}) 
$\tilde{\Phi}
 _{\text{EC}}(\hat{\rho}_{\text{CC}})$ does not depend upon such terms, 
 Eq.~(\ref{QCPA}) reduces to 
\begin{eqnarray} 
&& Q( \Phi_{\text{CC}}) \label{QCPA0}  =\max_{p\in [0,1]}  \Big\{ H_2(p)\\  \nonumber 
&& \qquad + 
 \max_{\tiny{\hat{\tau}_{\text{AA}}\in \mathfrak{S}_{{\text{A}}}}}
 \max_{ \tiny{
  \hat{\tau}_{\text{BB}}\in \mathfrak{S}_{{\text{B}}}}}
J_p(\Phi_{\text{AA}};\hat{\tau}_{\text{AA}}, \Phi_{\text{BB}};\hat{\tau}_{\text{BB}})
\Big\} \;. 
\end{eqnarray}
We can see that this expression involves an optimization only on the diagonal components of $\hat{\rho}_{\text{CC}}$.
The functional $J_p$ appearing in the above expression can be expressed as a rescaled convex combination of the
 coherent information terms of the channels $\Phi_{\text{AA}}$ and $\Phi_{\text{BB}}$. Explicitly 
\begin{eqnarray} 
 J_p &\equiv&
 \label{DEFJP} 
p I_{coh}(\Phi_{\text{AA}}; \hat{\tau}_{\text{AA}}) +  (1-p) I_{coh}(\Phi_{\text{BB}}; \hat{\tau}_{\text{BB}}) 
 \nonumber \\
 && - \Delta S_p (\tilde{\Phi}_{\text{EA}}[\hat{\tau}_{\text{AA}}], \tilde{\Phi}_{\text{EB}}[\hat{\tau}_{\text{BB}}])\; ,
\end{eqnarray} 
where for generic density matrices $\hat{\rho}'_{\text{EE}}$ and  $\hat{\rho}''_{\text{EE}}$ of E,  
we introduced  
\begin{eqnarray} 
 \Delta S_p (\hat{\rho}'_{\text{EE}},\hat{\rho}''_{\text{EE}})
  &\equiv& \nonumber   S\Big( p \hat{\rho}'_{\text{EE}} + (1-p)\hat{\rho}''_{\text{EE}}\Big) \\
&&- p S\left(  
\hat{\rho}'_{\text{EE}}\right)     - (1-p) S\left( \hat{\rho}''_{\text{EE}}\right) 
\;, \nonumber \\
\end{eqnarray} 
which is non-negative due to the concavity of the von Neumann entropy.
Notice that by simply specifying the above expression for the extreme cases  $p=1$ and $p=0$ one can
easily verify that Eq.~(\ref{QCPA0}) correctly complies with the bound in Eq.~(\ref{trivialbound}). 
On the contrary, an upper bound for $Q( \Phi_{\text{CC}})$ can be obtained 
 by dropping $\Delta S_p (\hat{\rho}'_{\text{EE}},\hat{\rho}''_{\text{EE}})$ in the right-hand-side
of Eq.~(\ref{DEFJP}), leading to the following inequality  
\begin{eqnarray}  
 Q( \Phi_{\text{CC}}) \label{QCPA0UP}  &\leq& \max_{p\in [0,1]}  \Big\{ H_2(p)  + p
Q( \Phi_{\text{AA}}) + (1-p) Q( \Phi_{\text{BB}}) \Big\}    \nonumber \\
&=& \log_2(2^{Q( \Phi_{\text{AA}})} +2^{Q( \Phi_{\text{BB}})} ) \;,
\end{eqnarray}
where we introduced  $Q(\Phi_{\text{AA}})$ and $Q(\Phi_{\text{BB}})$ using the optimization over $\hat{\tau}_{\text{AA}}$
and $\hat{\tau}_{\text{BB}}$, and where
 in the second line we carried out the maximization over $p$. This bound makes physical sense as
it implies that the dimension of the optimal noiseless subspace of $\Phi_{\text{CC}}$ cannot  be larger than
the direct sum of the noise-free subspace associated with the channels  $\Phi_{\text{AA}}$ and  $\Phi_{\text{BB}}$
when used independently. Notice also that the inequality~(\ref{QCPA0UP}) is saturated by taking $\Phi_{\text{CC}}$
to be the identity channel.

\subsection{Entanglement-assisted quantum capacity formula for  PCDS channels} \label{simplyEA} 
For the sake of completeness we report here the value of the entanglement assisted quantum capacity
 $Q_E(\Phi_{\text{XX}})$~\cite{ENT_ASS1,ENT_ASS2,QE} for the case of arbitrary (non-necessarily degradable) PCDS channels. 
We remind that if we  allow shared entanglement between sender and receiver the reliable transferring of quantum messages through the map $\Phi_{\text{XX}}$ can be improved via teleportation. The associated improvement is captured by the  following expression 
\begin{eqnarray}\label{eq:Ce=2Qe} 
Q_E(\Phi_{\text{XX}})&=&\frac{1}{2} \max_{\rho_{\text{XX}} \in \mathfrak{S}_{\text{X}}} I(\Phi_{\text{XX}}; \hat{\rho}_{\text{XX}})
\;, 
\end{eqnarray}
where now
\begin{eqnarray} \label{MUTUAL} 
I(\Phi_{\text{XX}}; \hat{\rho}_{\text{XX}}) &\equiv&   S( \hat{\rho}_{\text{XX}}) + I_{coh}(\Phi_{\text{XX}}; \hat{\rho}_{\text{XX}}) \;,
\end{eqnarray} 
 is the \textit{quantum mutual information}, which  being sub-additive  needs  no regularization even if the map $\Phi_{\text{XX}}$ is not degradable. 

In this case, besides Eq.~(\ref{ineq1}) we also invoke the  inequality 
\begin{eqnarray} 
S(\hat{\rho}_{\text{CC}})
\leq   p S\left(  
 \hat{\tau}_{\text{AA}}\right)   + (1-p) S\left( \hat{\tau}_{\text{BB}}\right) + H_2(p)\;, \nonumber \\
\end{eqnarray} 
that can be derived along the same line of reasoning. Replacing all this into Eq.~(\ref{eq:Ce=2Qe})
we get 
\begin{eqnarray} 
&& Q_E( \Phi_{\text{CC}}) \label{CEPA0}  =\max_{p\in [0,1]}  \Big\{ H_2(p)  \nonumber \\
&& \qquad + 
\frac{1}{2}  \max_{\tiny{\hat{\tau}_{\text{AA}}\in \mathfrak{S}_{{\text{A}}}}}
 \max_{ \tiny{
  \hat{\tau}_{\text{BB}}\in \mathfrak{S}_{{\text{B}}}}}
I_p(\Phi_{\text{AA}};\hat{\tau}_{\text{AA}}, \Phi_{\text{BB}};\hat{\tau}_{\text{BB}})
\Big\} \;, \nonumber \\
\end{eqnarray}
where now
\begin{eqnarray} 
 I_p&\equiv&  
 \label{DEFIP}   
 p I\left(  
 \Phi_{\text{AA}};\hat{\tau}_{\text{AA}}\right)     + (1-p) I\left( \Phi_{\text{BB}};\hat{\tau}_{\text{BB}}\right)  \nonumber \\ 
 && - \Delta S_p (\tilde{\Phi}_{\text{EA}}[\hat{\tau}_{\text{AA}}], \tilde{\Phi}_{\text{EB}}[\hat{\tau}_{\text{BB}}])\;, 
\end{eqnarray} 
with $I\left( \Phi_{\text{AA}};\hat{\tau}_{\text{AA}}\right)$ and $I\left( \Phi_{\text{BB}};\hat{\tau}_{\text{BB}}\right)$
the quantum mutual information functional of Eq.~(\ref{MUTUAL}) of $\Phi_{\text{AA}}$ and $\Phi_{\text{BB}}$ respectively. 
As in the case of the formula in Eq.~(\ref{QCPA0}) we can get a lower bound for it by taking $p=0,1$ and an upper bound by
dropping the term $\Delta S_p (\tilde{\Phi}_{\text{EA}}[\hat{\tau}_{\text{AA}}], \tilde{\Phi}_{\text{EB}}[\hat{\tau}_{\text{BB}}])$
in Eq.~(\ref{DEFIP}). This leads to the inequality
\begin{equation}
Q_E( \Phi^{(0)}_{\text{CC}}) \leq Q_E( \Phi_{\text{CC}}) \leq \log_2(2^{Q_E( \Phi_{\text{AA}})} +2^{Q_E( \Phi_{\text{BB}})} ) \;,
\end{equation} 
with $Q_E( \Phi^{(0)}_{\text{CC}})= \min\{ Q_E( \Phi^{(0)}_{\text{AA}}),Q_E( \Phi^{(0)}_{\text{BB}})\}$ as in Ref.~\cite{FUKUDA}.

\subsection{The special case of $\Phi_{\text{BB}}=\text{Id}_{\text{BB}}$} \label{SEC:ide} 

We now focus on the special case where the diagonal block $\Phi_{\text{BB}}$ of the PCDS channel $\Phi_{\text{CC}}$
defined in Eq.~(\ref{BRIEF}) 
corresponds to  the identity map $\text{Id}_{\text{BB}}$. Under this condition ${\cal H}_{\text{B}}$ is a decoherence-free subspace for the communication model. This implies that the value of 
$Q(\Phi_{\text{CC}})$ can always be lower bounded by $\log_2 d_{\text{B}}$, a condition that is automatically 
granted  by the inequality~(\ref{trivialbound}), noticing that in this case $Q( \Phi_{\text{BB}})=\log_2 d_{\text{B}}$.
Deeper insight on the model arises by observing that 
from Eq.~(\ref{COMPPHI}) we get 
 \begin{eqnarray}\label{ECCO1} 
\tilde{\Phi}_{\text{EB}}[{\hat{\Theta}}_{\text{BB}} ] = |0_{\text{E}}\rangle \! \langle 0_{\text{E}}| \; \mbox{Tr}_{\text{B}}[{\hat{\Theta}}_{\text{BB}} ]\;,  
\end{eqnarray} 
with $|0_{\text{E}}\rangle$ being an element of the orthonormal set $\{ |j_{\text{E}}\rangle\}_j$ of 
 ${\cal H}_{\text{E}}$. Accordingly from Eq.~(\ref{DEFJP})
we have 
\begin{eqnarray} 
 J_p&=& \nonumber 
 p S\left(  
  \Phi_{\text{AA}}[\hat{\tau}_{\text{AA}}]\right)     + (1-p) S\left(\hat{\tau}_{\text{BB}} \right) 
  \\  && - 
 S\Big( p \tilde{\Phi}
 _{\text{EA}}(\hat{\tau}_{\text{AA}})+ (1-p) |0_{\text{E}}\rangle \! \langle 0_{\text{E}}|
 \Big)  \nonumber \\
 &\leq&  p S\left(  
  \Phi_{\text{AA}}[\hat{\tau}_{\text{AA}}]\right)  + (1-p) \log_2 d_{\text{B}}  \nonumber 
   \\  && - 
 S\Big( p \tilde{\Phi}
 _{\text{EA}}(\hat{\tau}_{\text{AA}})+ (1-p) |0_{\text{E}}\rangle \! \langle 0_{\text{E}}|
 \Big) \;, 
\end{eqnarray} 
the upper bound being achieved by taking as input $\hat{\tau}_{\text{BB}}$  for B the completely mixed state $\hat{P}_{\text{BB}}/d_{\text{B}}$. 
Hence the capacity formulas in Eqs.~(\ref{QCPA0}) and (\ref{CEPA0}) now  write respectively 
\begin{eqnarray}\label{eq:quantum capacity double maxim}
Q(\Phi_{\text{CC}})  &=& \max_{ p\in[0,1]}\Big\{ H_2(p)+ (1-p) \log_2 d_{\text{B}} \nonumber  \\
  && + \max_{\hat{\tau}_{\text{AA}} \in  \mathfrak{S}_{{\text{A}}} }  \Big\{  p S\left(  
  \Phi_{\text{AA}}[\hat{\tau}_{\text{AA}}]\right) \label{QB}   \nonumber  \\  \nonumber 
  && - S(p\tilde{\Phi}_{\text{EA}}[ \hat{\tau}_{\text{AA}}]+  (1-p)  |0_{\text{E}}\rangle \! \langle 0_{\text{E}}| )  \Big\}\Big\}\;, \nonumber \\
\end{eqnarray} 
which holds true for all choices of CPT maps $\Phi_{\text{AA}}$ that are degradable, and 
\begin{eqnarray}\label{eq:quantum capacity double maximCE}
Q_E(\Phi_{\text{CC}})  &=& \max_{ p\in[0,1]}\Big\{ H_2(p) + {(1-p)} \log_2 d_{\text{B}}  \nonumber  \\
  && +\frac{1}{2}  \max_{\hat{\tau}_{\text{AA}} \in  \mathfrak{S}_{{\text{A}}} }  \Big\{  p S\left(  
  \hat{\tau}_{\text{AA}}\right)  + p S\left(  
  \Phi_{\text{AA}}[\hat{\tau}_{\text{AA}}]\right) \label{QB}  \nonumber  \\ \nonumber 
  && - S(p\tilde{\Phi}_{\text{EA}}[ \hat{\tau}_{\text{AA}}]+  (1-p)  |0_{\text{E}}\rangle \! \langle 0_{\text{E}}| )  \Big\}\Big\}\;, \nonumber \\
\end{eqnarray} 
that instead applies also for non degradable CPT maps $\Phi_{\text{AA}}$ -- both expressions 
  now involving only an optimization with respect to $\hat{\tau}_{\text{AA}}$ and $p$. 

Notice that the relatively simple expression reported in Eq.~(\ref{eq:quantum capacity double maxim}) 
paves the way to refine a little the
 lower  bound discussed in Sec.~\ref{SEC:COMP} for general PCDS channels. 
 In particular, assume that there exists a density matrix $\hat{\rho}^*_{\text{AA}}$ of A such that 
  the complementary channel $\tilde{\Phi}_{\text{EA}}$ of  $\tilde{\Phi}_{\text{AA}}$
fulfills the following identity 
\begin{eqnarray}\label{FFd1}  \tilde{\Phi}_{\text{EA}}[  \hat{\rho}^*_{\text{AA}}]=  |0_{\text{E}}\rangle \! \langle 0_{\text{E}}|\;,
\end{eqnarray} 
with $|0_E\rangle$ being the pure vector that via Eq.~(\ref{ECCO1}) defines the action of $\tilde{\Phi}_{\text{EB}}$. Interestingly enough, in Appendix~\ref{appPURE}  we show  that this special requirement can always be met if the channel  
${\Phi}_{\text{AA}}$ admits a fixed point state that is pure (examples of those maps are  provided by the cases
studied in  Sec.~\ref{application}  and 
Sec.~\ref{application1}). 
Under the hypothesis in Eq.~(\ref{FFd1}),  setting $\hat{\tau}_{\text{AA}} = \hat{\rho}^*_{\text{AA}}$ in the right-end-side of 
Eqs.~(\ref{eq:quantum capacity double maxim}) and dropping a positive term we can then arrive to the inequality 
\begin{eqnarray}
Q(\Phi_{\text{CC}}) 
  &\geq&  \max_{ p\in[0,1]}\Big\{H_2(p) +  (1-p) \log_2 d_{\text{B}}  
   \Big\}\nonumber \\
  &=& \log_2 (d_{\text{B}}+1)\;. \label{NEWBOUND} 
\end{eqnarray} 
For $ \log_2 (d_{\text{B}}+1)> Q(\Phi_{\text{AA}})$ represents an improvement with respect to the  general lower bound given in Eq.~(\ref{trivialbound}). 
At the physical level Eq.~(\ref{NEWBOUND}) implies that under the condition in Eq.~(\ref{FFd1}) the model admits the presence of a decoherence-free subspace.  The dimension of this subspace is $d_{\text{B}}+1$ and is slightly larger than the value  $d_{\text{B}}$ 
that is granted for free by having the block B preserved during the evolution. 
An interesting consequence of Eq.~(\ref{NEWBOUND}) can finally be
drawn by comparing it with Eq.~(\ref{QCPA0UP}). Indeed in the present case, due to the fact that $Q( \Phi_{\text{BB}}=\text{Id}_{\text{BB}})=\log_2 d_{\text{B}}$, such an upper bound reduces to
\begin{eqnarray}  
 Q( \Phi_{\text{CC}}) \label{QCPA0UPnew}  \leq  \log_2(2^{Q( \Phi_{\text{AA}})} +d_{\text{B}}) \;,
\end{eqnarray}
whose right-hand-side term exactly matches that of the lower bound of Eq.~(\ref{NEWBOUND}) whenever  $Q( \Phi_{\text{AA}})=0$.
Putting all this together we can  then arrive to the following observation 
 \begin{lemma} 
Let $\Phi_{\text{CC}}$ be  a PCDS quantum channel~(\ref{eq:channel decomposition}) 
with $\Phi_{\text{BB}}=\text{Id}_{\text{BB}}$. If  $\Phi_{\text{AA}}$ is a zero-capacity (i.e. $Q( \Phi_{\text{AA}})=0$),
degradable map admitting a pure fixed point state then we have 
\begin{eqnarray}  
 Q( \Phi_{\text{CC}}) \label{QCPA0UPnewNEW}  = \log_2(d_{\text{B}}+1) \;.
\end{eqnarray} \label{lemma1}
\end{lemma}
Explicit examples of $\Phi_{\text{CC}}$ obeying the 
structural constraints imposed by the Lemma will be presented in  Secs.~\ref{application} and~\ref{application1},
 together with a rather important consequence of the identity in Eq.~(\ref{QCPA0UPnewNEW}).

\section{Applications}  \label{Sec:application}

Here we report few applications of the identity in Eq.~(\ref{eq:quantum capacity double maxim}) 
 that allows us to 
fully characterize the quantum capacity of a large class of nontrivial PCDS quantum channels, including some
specific examples of CPT maps which are not degradable. 

\subsection{Purely Dephasing channels} \label{Sec:PURELYDEP} 
\begin{figure}[h!]
  \includegraphics[width=\linewidth]{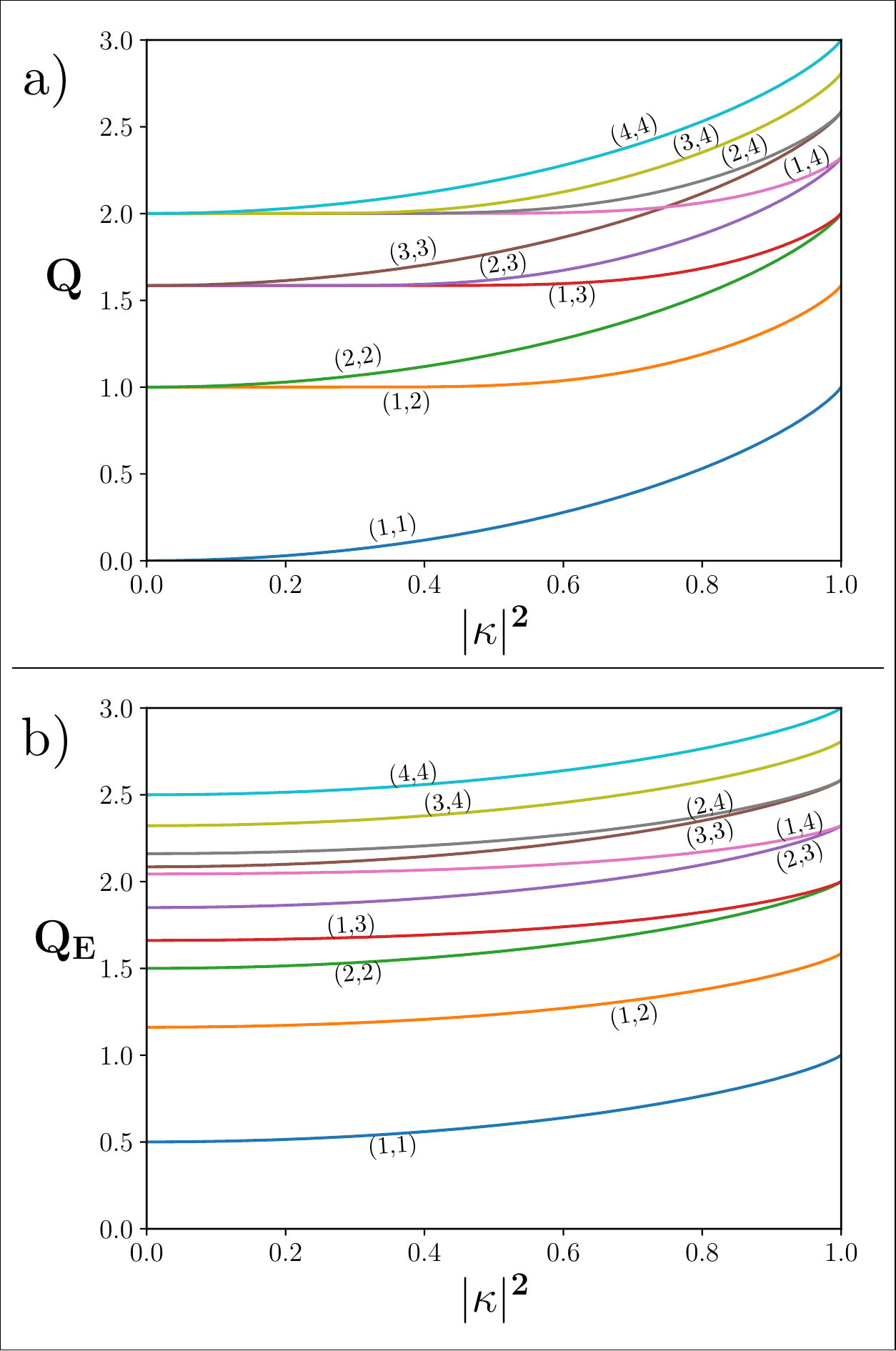}
  \caption{\textbf{a)} Quantum capacity $Q$ of the purely dephasing channel $\Delta^{(\kappa)}_{\text{CC}}$
  of Eq.~(\ref{D2def})  for some values of the couple $(d_{\text{A}},d_{\text{B}})$ w.r.t. the dephasing parameter $|\kappa|^2$. For $d_{\text{A}}=d_{\text{B}}=1$ we recover the quantum capacity of the qubit dephasing channel of \cite{DEVSHOR}.
 \textbf{b)} Entanglement assisted quantum capacity  $Q_E$ of $\Delta^{(\kappa)}_{\text{CC}}$ for some values of the couple $(d_{\text{A}},d_{\text{B}})$ w.r.t. the dephasing parameter $|\kappa|^2$.
It is worth observing that the curves associated with the same value of the ratio $d_{\text{A}}/d_{\text{B}}$ 
differs only by an additive constant as predicted in the main text, and that the presence of the entanglement resource removes the degeneracy of
the $Q(\Delta^{(\kappa)}_{\text{CC}})$ capacity for $\kappa=0$. The monotonic behavior of the plotted curves follows from the channel DPI and from the trivial composition rules obeyed by  
the maps $\Delta^{(\kappa)}_{\text{CC}}$.}
  \label{fig:dephasing}
\end{figure}

As a first example of PCDS channels $\Phi_{\text{CC}}$ described in Sec.~\ref{SEC:ide} 
 we focus on the  purely dephasing maps \cite{DEPH_MEMO, DEPHASING}  $\Delta^{(\kappa)}_{\text{CC}}$ of Eq.~(\ref{D2def}).
 Accordingly in this case both 
$\Phi_{\text{BB}}$  and $\Phi_{\text{AA}}$ are the identity transformation and 
 we can assign the Kraus set of the model by taking the following operators 
 \begin{eqnarray} \label{KRAUSPURE} 
 {\hat{M}}_{\text{CC}}^{(0)} =   \kappa {\hat{P}}_{\text{AA}}+  {\hat{P}}_{\text{BB}} \;, \quad 
  {\hat{M}}_{\text{CC}}^{(1)} = \sqrt{1- |\kappa|^2}  {\hat{P}}_{\text{AA}}\;.\nonumber \\
 \end{eqnarray}
Via Eq.~(\ref{COMPPHI}) this leads us to Eq.~(\ref{ECCO1}) for the complementary channel 
$\tilde{\Phi}_{\text{EB}}$ 
and to
\begin{eqnarray}\label{ECCO2} 
\tilde{\Phi}_{\text{EA}}[{\hat{\Theta}}_{\text{AA}} ] = |v^{(\kappa)}_{\text{E}}\rangle \langle v^{(\kappa)}_{\text{E}}| \; \mbox{Tr}_{\text{A}}[{\hat{\Theta}}_{\text{AA}} ]\;,  
\end{eqnarray} 
where now $|v_{\text{E}}\rangle$ is the pure state vector 
\begin{eqnarray}
|v^{(\kappa)}_{\text{E}}\rangle \equiv \kappa |0_{\text{E}}\rangle + \sqrt{1 - |\kappa|^2} |1_{\text{E}}\rangle \label{defv} \;. 
\end{eqnarray} 
Since in the present case $\Phi_{\text{AA}}$ is the identity channel, hence degradable, we can compute the quantum capacity of 
$\Delta^{(\kappa)}_{\text{CC}}$  via the single letter formula in Eq.~(\ref{eq:quantum capacity double maxim}) which, 
 by trivially upper-bounding $S\left(  
  \Phi_{\text{AA}}[\hat{\tau}_{\text{AA}}]\right)$ with $\log_2 d_{\text{A}}$, 
rewrites as   
\begin{eqnarray}\nonumber 
Q(\Delta^{(\kappa)}_{\text{CC}})  &=& \max_{ p\in[0,1]}\Big\{ H_2(p)  + p  \log_2 d_{\text{A}}  +  (1-p) \log_2 d_{\text{B}}   \\ 
  && - S(p |v^{(\kappa)}_{\text{E}}\rangle \! \langle v^{(\kappa)}_{\text{E}}|+(1-p)|0_{\text{E}}\rangle \! \langle 0_{\text{E}}|   ) \Big\}\nonumber\\
  &=&\log_2 d_{\text{B}} +   \max_{ p\in[0,1]} \Big\{  H_2(p)+p\log_2(d_{\text{A}}/d_{\text{B}}) 
  \nonumber \\
  &&- H_2\left( \tfrac{1+\sqrt{1-4p(1-p)(1-|\kappa|^2)}
  }{2}\right) 
  \Big\} \label{eq:quantum capacity double maximDEF}\;.
  \end{eqnarray} 
In the limiting cases $|\kappa|=1$ (no noise) and $\kappa=0$ (full dephasing)
the maximization can be explicitly performed leading to the expected results of Eqs.~(\ref{QK1}) and (\ref{QK0}),
respectively. For all the other choices of $\kappa$ we resort to numerical evaluation and report the obtained results  in 
Fig.~\ref{fig:dephasing} a). Partial analytical information can however be recovered by noticing that the function we have to optimize with respect  to $p$ depends, apart from  the noise coefficient $|\kappa|$, 
  only upon the ratio $d_{\text{A}}/d_{\text{B}}$. From this fact, by simple analytical considerations it follows  that functions~$Q(\Delta^{(\kappa)}_{\text{CC}})$  associated with models with same value of ratio
  $d_{\text{A}}/d_{\text{B}}$ will only differ by an additive constant. 
  Furthermore, in the special case where 
 $d_{\text{A}}/d_{\text{B}}=1$
 the maximization
 can be again carried out analytically, e.g. by noticing that the associated functional is symmetric for exchange of $p$ and $1-p$:
 accordingly we can conclude that in this case the optimal value for $p$ is $1/2$, implying  
\begin{eqnarray} \label{DEFQ} 
\hspace{-15pt}Q(\Delta^{(\kappa)}_{\text{CC}})&=&1-H_2((1-|\kappa |^2)/2)+\log_2(d_\text{A})\;,\end{eqnarray} 
For $d_{\text{A}}=1$ this expression correctly reproduces the capacity formula of Ref.~\cite{DEVSHOR}
for the qubit ($d_{\text{C}}=2)$ dephasing channel. It's worth noticing from Fig.~\ref{fig:dephasing} a) that depending on the combination of $(d_{\text{A}},d_{\text{B}})$ a structure among the channels emerges. The noiseless subspace associated with $d_{\text{B}}$ defines a ``multiplet'' of curves that converge to $\log_2(d_{\text{B}})$ at $\kappa \sim 0$ and spread with increasing $\kappa$ toward the values $\log_2(d_{\text{A}}+d_{\text{B}})$, never intersecting each other. Intersections can take place between elements of different multiplets, as happens e.g. for the curves (3,3) and (1,4). In this case we can see that when $\kappa \gtrsim 0.75$, having 3 decohering levels and 3 noiseless performs better than having only 1 decohering level and 4 noiseless.

 Similar conclusions can be drawn for the entanglement assisted capacity  of $\Delta^{(\kappa)}_{\text{CC}}$, which 
 from Eq.~(\ref{eq:quantum capacity double maximCE})  we express as 
  \begin{eqnarray}  Q_E(\Delta^{(\kappa)}_{\text{CC}})  &=& \max_{ p\in[0,1]}\Big\{ H_2(p)  + {p} \log_2 d_{\text{A}}  +  {(1-p)} \log_2 d_{\text{B}}  \nonumber \\ 
  && - \frac{1}{2} S(p |v^{(\kappa)}_{\text{E}}\rangle \! \langle v^{(\kappa)}_{\text{E}}|+(1-p) |0_{\text{E}}\rangle \! \langle 0_{\text{E}}|  ) \Big\} \nonumber \\ 
  \nonumber\\
  &=&\log_2 d_{\text{B}} +   \max_{ p\in[0,1]} \Big\{  H_2(p)+p\log_2(d_{\text{A}}/d_{\text{B}}) 
  \nonumber \\
  &&- \frac{1}{2}H_2\left( \tfrac{1+\sqrt{1-4p(1-p)(1-|\kappa|^2)}
  }{2}\right) 
  \Big\} \label{eq:quantum capacity double maximDEF1}\;,
  \end{eqnarray} 
whose values are plotted in Fig.~\ref{fig:dephasing} b) (notice again that 
for $d_{\text{A}}= d_{\text{B}}$ the optimization can be performed analytically resulting in 
$Q_E(\Delta^{(\kappa)}_{\text{CC}})=1-\frac{1}{2} H_2((1-|\kappa|^2)/2)+\log_2(d_\text{A})$).

\subsection{Multi-level Amplitude Damping channels } \label{application} 

\begin{figure}[t!]
  \includegraphics[width=0.75\linewidth]{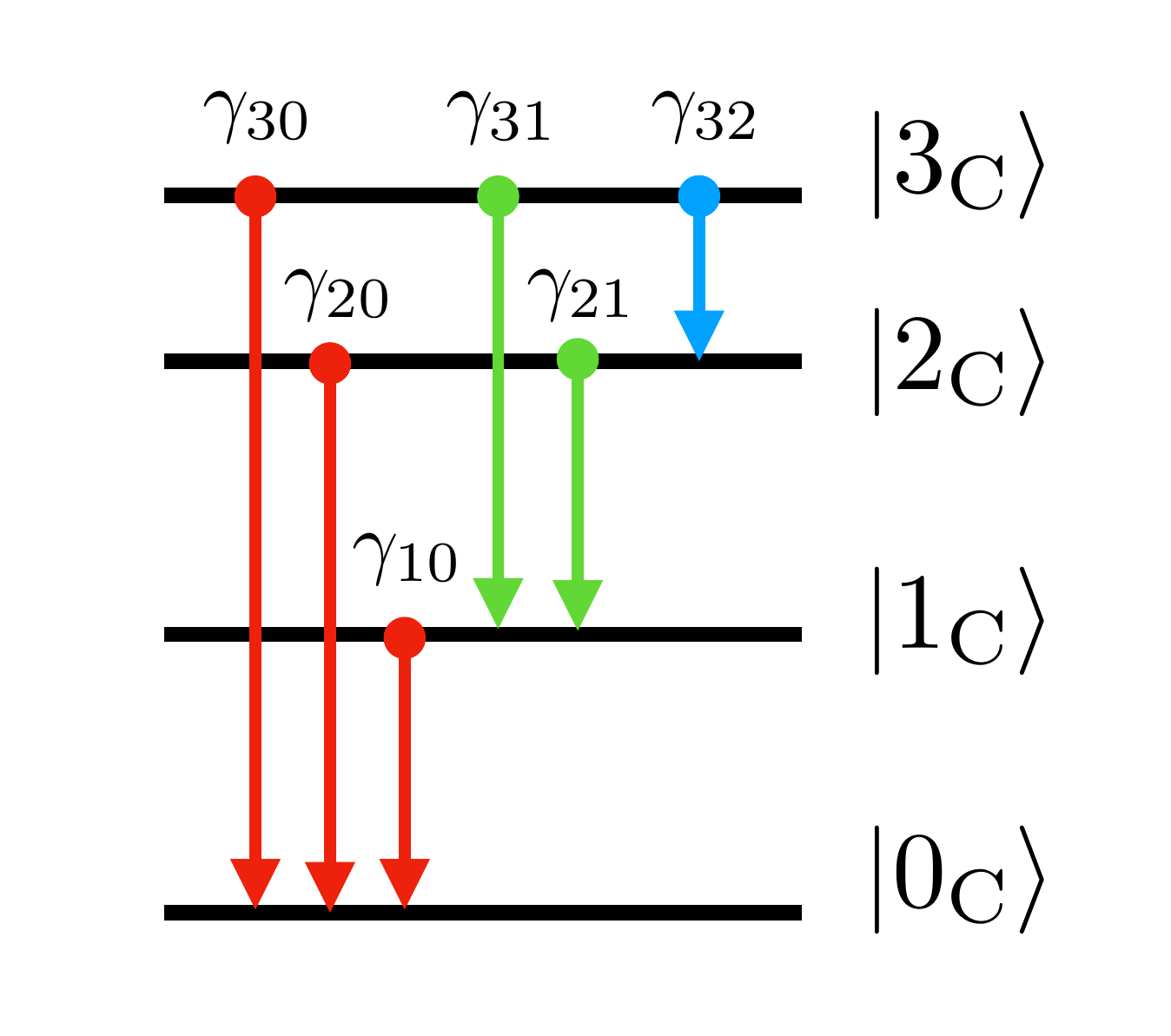}
  \caption{Schematic representation of a
MAD channel acting on a system C of dimension $d_{\text{C}}=4$: each arrow represents a decaying process where given 
$j>i$, 
the upper level $|j_{\text{C}}\rangle$ tends to relax toward the lower level $|i_{\text{C}}\rangle$ at rate $\gamma_{ji}$. Notice that by construction the ground state $|0_\text{C}\rangle$ is a fixed point of the evolution. 
An example of a PCDS map can be obtained for instance by imposing $\gamma_{30}=\gamma_{31}=\gamma_{31}=\gamma_{21}=0$ (in this case A and B are both bi-dimensional subsets spanned by the vectors
$\ket{0_{\text{C}}}$, $\ket{1_{\text{C}}}$  and  $\ket{2_{\text{C}}}$, $\ket{3_{\text{C}}}$, respectively. The single non-zero decay rate
 MAD channel ${\Omega}_{\text{CC}}^{[\gamma]}$  is finally obtained by taking $\gamma_{10}=\gamma$ and setting all the other rates equal to zero: notice that in this case restricting the input states to the $3$-dimensional subspace 
 spanned by $|0_{\text{C}}\rangle$, $|2_{\text{C}}\rangle$, and $|3_{\text{C}}\rangle$, they will be preserved by the action of the noise. }
  \label{fig:ADC}
\end{figure}

As a second example we now focus on a multi-level version of the qubit Amplitude Damping channel~\cite{QUBITADC}, hereafter indicated as MAD channels in brief, which describes the probability for levels of a $d_{\text{C}}$-dimensional system to decay into each other~\cite{QUTRIT_ADC}. 
In their most general form, given $\{|i_{\text{C}}\rangle\}_{i=0, \cdots, d_{\text{C}}-1}$ an orthornormal basis for 
${\cal H}_{\text{C}}$, 
these maps can be assigned by introducing the set of
Kraus operators $\{\hat{M}_{\text{CC}}^{(0)}\} \bigcup \{ \hat{M}_{\text{CC}}^{(ij)} \}_{i<j}$  formed by the $d_{\text{C}}(d_{\text{C}}-1)/2$ matrices  \begin{eqnarray}\label{eq.Kraus}
\hat{M}_{\text{CC}}^{(ij)}&\equiv&\sqrt{\gamma_{ji}}\ket{i_{\text{C}}}\!\!\bra{j_{\text{C}}}, \qquad \forall i<j \;, \end{eqnarray} 
with  $\gamma_{ji}$ real quantities on the interval $[0,1]$ describing the decay rate from the $j$-th to the $i$-th level (see Fig.~\ref{fig:ADC}). The damping parameters fulfill the conditions
\begin{eqnarray} \label{BIGMA} 
\xi_j\equiv \sum_{0\leq i <j} \gamma_{ji} \leq 1\;,  \qquad \forall j= 1,\cdots, d_{\text{C}}-1\;.
\end{eqnarray} 
The Kraus set is completed by 
\begin{eqnarray}
\hspace{-10pt}\hat{M}_{\text{CC}}^{(0)} &\equiv&  \ket{0_{\text{C}}}\!\!\bra{0_{\text{C}}} + 
\sum\limits_{1\leq j\leq d_{\text{C}}-1} \sqrt{1 -\xi_j} \ket{j_{\text{C}}}\!\!\bra{j_{\text{C}}}\;.
\end{eqnarray}
  Besides providing effective  description of the noisy evolution of energy dissipation of atomic models, MAD channels have a rather rich structure. 
  Limit cases are those where all the $\gamma_{ji}$ are zero, corresponding to the identity channel $\text{Id}_{\text{CC}}$, and the cases where equality holds in Eq.~(\ref{BIGMA}) leaving the level $j$ totally depopulated. Most importantly for us, by properly tailoring the values of the parameters $\gamma_{ji}$,
 MAD channels can be used to construct nontrivial examples of PCDS channels. This happens, for instance, whenever the set of  rates which are explicitly non zero can be split into two distinct groups of  $\gamma_{ji}$ characterized by values of the indexes $j,i$ which span disjoint sets -- see caption of Fig.~\ref{fig:ADC}. 
For the purpose of the present analysis we shall focus  
on the  special class of these  channels characterized by a single non-zero decay rate~\cite{QUTRIT_ADC}.
Without loss of generality we choose the not null decaying parameter $\gamma \in [0,1]$ to be the one connecting levels $\ket{0_{\text{C}}}$ and $\ket{1_{\text{C}}}$. We'll indicate then this channel as
${\Omega}_{\text{CC}}^{[\gamma]}$. Under this condition the Kraus set contains only 
two terms 
  \begin{eqnarray}\label{eq.Kraus11}
\hat{M}_{\text{CC}}^{(01)}&\equiv&\sqrt{\gamma}\ket{0_{\text{C}}}\!\!\bra{1_{\text{C}}} \;, \nonumber \\
\hat{M}_{\text{CC}}^{(0)} &\equiv&  \ket{0_{\text{C}}}\!\!\bra{0_{\text{C}}} + 
 \sqrt{1 -\gamma} \ket{1_{\text{C}}}\!\!\bra{1_{\text{C}}}  \nonumber
 + \sum\limits_{2\leq j\leq d_{\text{C}}-1} \ket{j_{\text{C}}}\!\!\bra{j_{\text{C}}}\; , \nonumber \\
 \end{eqnarray} 
which can be easily cast in the PCDS canonical form of Theorem~\ref{thm:necandsuff}. This is done by 
identifying ${\cal H}_{\text{A}}$  with the bi-dimensional ($d_{\text{A}}=2$) subset spanned by the vectors $\ket{0_{\text{A}}}\equiv \ket{0_{\text{C}}}$, $\ket{1_{\text{A}}}\equiv \ket{1_{\text{C}}}$, and ${\cal H}_{\text{B}}$
with the Hilbert space of dimension $d_{\text{B}}=d_{\text{C}}-2$ spanned by the vectors $\{|i_{\text{B}}\rangle \equiv  |(i+2)_{\text{C}}\rangle\}_{i=0,\cdots, d_{\text{B}}-1}$.
Accordingly ${\Omega}_{\text{CC}}^{[\gamma]}$ can be expressed as in Eq.~(\ref{eq:channel decomposition}) 
with the diagonal terms  given respectively by the identity map ${\text{Id}}_{\text{BB}}$ on B, and by the standard qubit Amplitude Damping Channel (ADC)  ${\Omega}_{\text{AA}}^{[\gamma]}$, described by the Kraus elements 
$\hat{M}_{\text{AA}}^{(01)}\equiv\sqrt{\gamma}\ket{0_{\text{A}}}\!\!\bra{1_{\text{A}}}$ 
$\hat{M}_{\text{AA}}^{(0)}\equiv\ket{0_{\text{A}}}\!\!\bra{0_{\text{A}}} + (1-\gamma)\ket{1_{\text{A}}}\!\!\bra{1_{\text{A}}}$.
Notice also that  any even value of $d_{\text{C}}$ can be seen as the dimension of a tensor Hilbert space  $\mathcal{H}_{\text{C}_1}\otimes \mathcal{H}_{\text{C}_2}$ s.t. $d_{\text{C}_1} d_{\text{C}_2} =d_{\text{C}}$. We can then see the MAD
channel ${\Omega}_{\text{CC}}^{[\gamma]}$ 
as a fully correlated ADC on $\mathcal{H}_{\text{C}_1}\otimes \mathcal{H}_{\text{C}_2}$ analogous to those studied by D'Arrigo et al. in Ref.~\cite{DARRIGO}. There they studied the case for which $d_{\text{C}_1} =d_{\text{C}_2} =2$, that damps 
the 2-qubits state $\ket{11}$ in $\ket{00}$ and leaves the subspace spanned by $\ket{01}$ and $\ket{10}$ untouched. 

We now proceed with the explicit evaluation of the quantum capacity of ${\Omega}_{\text{CC}}^{[\gamma]}$. 
As a preliminary observation we establish two facts that hold true for the entire spectrum of the values of the parameter $\gamma$.
First of all, as in the case of their qubit counterpart ${\Omega}_{\text{AA}}$, the set of MAD channel ${\Omega}_{\text{CC}}$ is closed under 
channel composition. In particular 
given $\gamma_1,\gamma_2\in[0,1]$, we have 
${\Omega}_{\text{CC}}^{[\gamma_1]}\circ {\Omega}_{\text{CC}}^{[\gamma_2]}={\Omega}_{\text{CC}}^{[\gamma_3]}$
with $\gamma_3\equiv \gamma_1 + \gamma_2 -\gamma_1\gamma_2$.
Noticing that $\gamma_3$ is larger than $\gamma_1$ and $\gamma_2$, we can hence invoke the coherent information DPI to 
establish that  $Q({\Omega}_{\text{CC}}^{[\gamma]})$  must be monotonically decreasing w.r.t. $\gamma$, i.e. 
\begin{eqnarray}  \label{MON} 
Q({\Omega}_{\text{CC}}^{[\gamma]}) \geq Q({\Omega}_{\text{CC}}^{[\gamma']}) \qquad \forall \gamma\leq \gamma'\;. 
\end{eqnarray} 
Second we notice that for all  $\gamma$ values we have that the 
$d_{\text{C}}-1$ dimensional subspace ${\cal H}'_{\text{C}}$, spanned by all the  vectors of the basis  $\{|i_{\text{C}}\rangle\}_{i=0, \cdots, d_{\text{C}}-1}$  but $|1_{\text{C}}\rangle$,  is fully preserved by the action of ${\Omega}_{\text{CC}}^{[\gamma]}$, i.e. ${\Omega}_{\text{CC}}^{[\gamma]}[\hat{\rho}_{\text{CC}}] = \hat{\rho}_{\text{CC}}$ $\forall$ $\hat{\rho}_{\text{CC}}\in \mathfrak{S}({\cal H}'_{\text{C}})$.
Accordingly the model allows for the reliable transfer of at least $\log_2 (d_{\text{C}}-1)$ qubits, leading to the  following inequality 
  \begin{eqnarray} 
 Q({\Omega}_{\text{CC}}^{[\gamma]}) \geq \log_2 (d_{\text{C}}-1)=  \log_2 (d_{\text{B}}+1)\;,\label{lowerbbb} 
 \end{eqnarray} 
 which subsides  the lower bound  $Q({\Omega}_{\text{CC}}^{[\gamma]}) \geq \log_2 d_{\text{B}}$ that follows
 from Eq.~(\ref{DSCAP}). 

\begin{figure}[t!]
  \includegraphics[width=\linewidth]{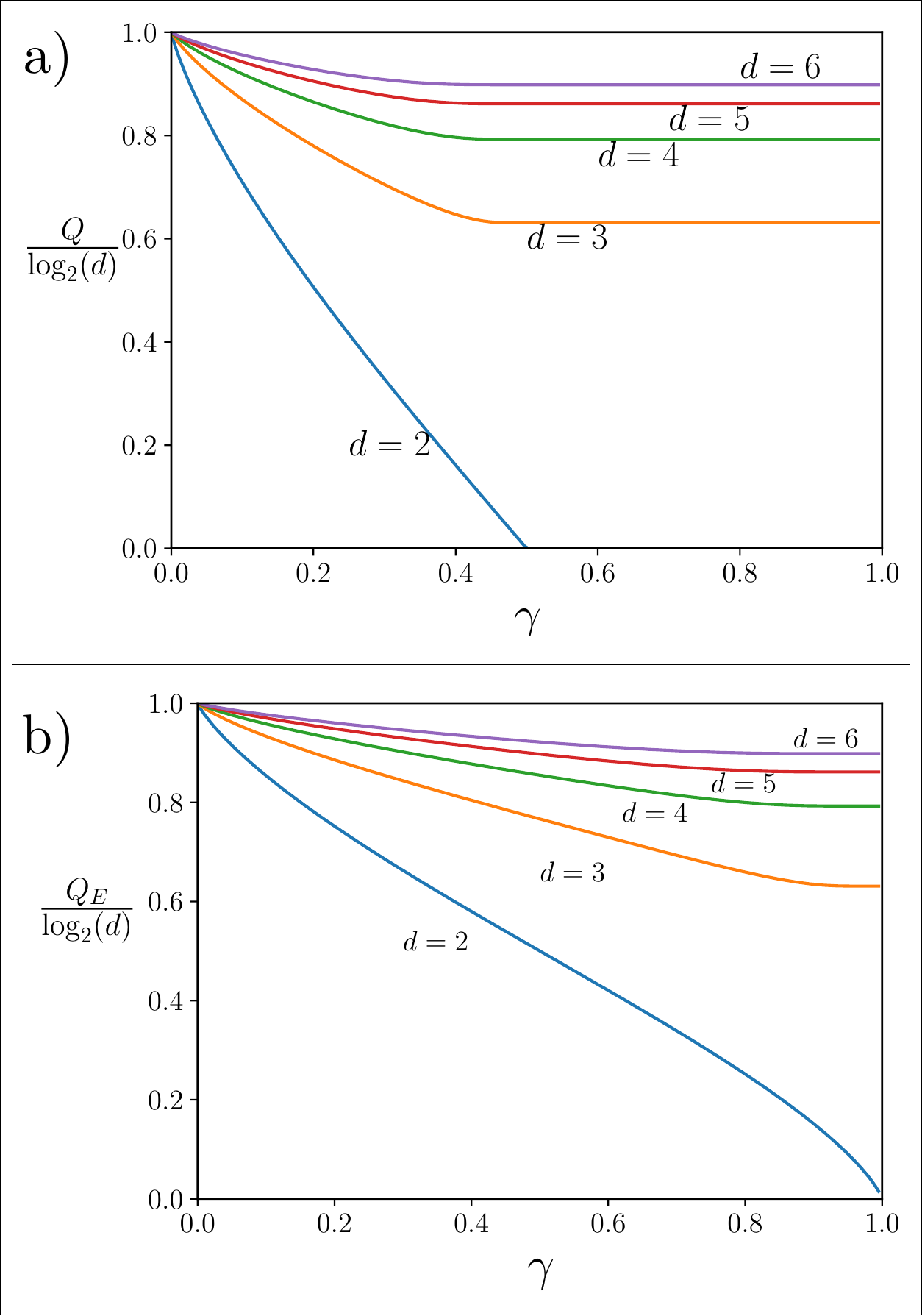}
  \caption{Normalized quantum capacities \textbf{a)}  and entanglement assisted quantum capacities   \textbf{b)}  of single decay MAD channel ${\Omega}_{\text{CC}}^{[\gamma]}$ 
   for various dimensions $d_{\text{C}}$. Notice that the $d_{\text{C}}=2$ case corresponds to the qubit ADC \cite{QUBITADC} and the $d_{\text{C}}=4$ case to the fully correlated ADC of~\cite{DARRIGO}. The channel is degradable only for 
    $\gamma\leq 1/2$; for higher values of the rate the quantum capacity is constant and equal to $\log_2(d_{\text{B}}+1)$, see Eq.~(\ref{EXALOW}). The non increasing functional dependence of  $Q({\Omega}_{\text{CC}}^{[\gamma]})$ and $Q_E({\Omega}_{\text{CC}}^{[\gamma]})$ upon   $\gamma$ is a consequence  of the composition rule of the MAP channels and by the channel
    DPI.
}
  \label{fig:quditADC}
\end{figure}

Let's then proceed with the explicit evaluation of the capacity. 
To begin with, we remind that the qubit ADC  ${\Omega}_{\text{AA}}^{[\gamma]}$ 
 is known  to
  be degradable for $0\leq \gamma \leq 1/2$ and antidegradable for $1/2\leq  \gamma\leq 1$~\cite{QUBITADC}. Invoking hence Theorem~\ref{thm:optimalDeltaM} we can conclude that the MAD channel 
 ${\Omega}_{\text{CC}}^{[\gamma]}$ is degradable if and only if  $0\leq \gamma \leq 1/2$.
 For this values (and only for those values) we can hence compute  $Q({\Omega}_{\text{CC}}^{[\gamma]})$  with the single letter formula in Eq.~(\ref{eq:quantum capacity double maxim}).
  Specifically, remembering that 
the complementary channel of the qubit ADC ${\Omega}_{\text{AA}}^{[\gamma]}$ for given $\gamma$ is unitarily equivalent to the qubit ADC ${\Omega}_{\text{AA}}^{[1-\gamma]}$~\cite{QUBITADC}, we can write 
\begin{eqnarray}\label{eq:quantum capacity double maxim323}
Q({\Omega}_{\text{CC}}^{[\gamma]})  &=& \max_{ p\in[0,1]}\Big\{ H_2(p)+ (1-p) \log_2 d_{\text{B}} \nonumber  \\
  && + \max_{\hat{\tau}_{\text{AA}} \in  \mathfrak{S}_{{\text{A}}} }  \Big\{  p S\left(  
 {\Omega}_{\text{AA}}^{[\gamma]}[\hat{\tau}_{\text{AA}}]\right) \label{QB}   \nonumber  \\  \nonumber 
  && - S(p{\Omega}_{\text{AA}}^{[1-\gamma]}[ \hat{\tau}_{\text{AA}}]+  (1-p)  |0_\text{A}\rangle \!\langle 0_\text{A}| )  \Big\}\Big\}\;, \nonumber \\
\end{eqnarray} 
where without loss of generality we identified the vector $|0_{\text{E}}\rangle$ of the environment E with the  ground state
$|0_{\text{A}}\rangle$ of~A. A numerical  evaluation of this function is reported  in Fig.~\ref{fig:quditADC} a) for different choices of $d_{\text{B}}$. Notice  in particular that for $\gamma=1/2$ we get 
\begin{eqnarray} \label{eta12}  Q({\Omega}_{\text{CC}}^{[1/2]})= \log_2(d_{\text{B}}+1),\end{eqnarray}  something that can be analytically proven as a direct consequence of Lemma~\ref{lemma1}. This is due to the fact that in this case 
 $Q({\Omega}_{\text{AA}}^{[\gamma=1/2]})=0$ (the channel ${\Omega}_{\text{AA}}^{[\gamma=1/2]}$ being
 both degradable and antidegradable), and 
 ${\Omega}_{\text{AA}}^{[\gamma=1/2]}$ admits the pure state $|0_\text{A}\rangle$  as  fixed point~\cite{QUBITADC}, i.e. 
 ${\Omega}_{\text{AA}}^{[\gamma=1/2]}[|0_{\text{A}}\rangle\! \langle 0_{\text{A}}|]=|0_{\text{A}}\rangle \! \langle 0_{\text{A}}|$.
 
 What about the capacity of ${\Omega}_{\text{CC}}^{[\gamma]}$ for $\gamma>1/2$? In this case Eq.~(\ref{eq:quantum capacity double maxim323}) does not necessarily apply due to the fact that ${\Omega}_{\text{CC}}^{[\gamma]}$ is provably not 
 degradable. Observe that in this regime, at variance with its qubit counterpart ${\Omega}_{\text{AA}}^{[\gamma]}$,
 ${\Omega}_{\text{CC}}^{[\gamma]}$  is  also certainly non antidegradable as a trivial consequence of 
 the bound in Eq.~(\ref{lowerbbb}) which prevents the  quantum capacity from being zero. 
 Accordingly the explicit evaluation of $Q({\Omega}_{\text{CC}}^{[\gamma]})$ for $\gamma>1/2$ 
 would require in principle to pass through the cumbersome regularization of Eq.~(\ref{DEFQUA}). It turns out however that in this case  we can explicitly compute $Q({\Omega}_{\text{CC}}^{[\gamma]})$ showing that it must keep the constant value it achieved
for $\gamma=1/2$, i.e. 
\begin{eqnarray}Q({\Omega}_{\text{CC}}^{[\gamma]})= \log_2(d_{\text{B}}+1), \qquad \forall \gamma \in [1/2,1]\;. 
\end{eqnarray} 
This indeed follows from  Eq.~(\ref{eta12}), the monotonicity condition in Eq.~(\ref{MON}), and the lower bound in Eq.~(\ref{lowerbbb}) which together imply 
\begin{eqnarray} \label{EXALOW} 
Q({\Omega}_{\text{CC}}^{[1/2]}) \geq Q({\Omega}_{\text{CC}}^{[\gamma]})  \geq \log_2(d_{\text{B}}+1)\;. 
\end{eqnarray}
It follows then, even if the channel is not degradable for $1/2< \gamma\leq 1$, that $Q({\Omega}_{\text{CC}}^{[\gamma]})=Q^{(1)}({\Omega}_{\text{CC}}^{[\gamma]})$ $\forall \gamma$.

All these results have been summarized in Fig.~\ref{fig:quditADC} a). In Fig.~\ref{fig:quditADC} b) instead we report the value
of $Q_E({\Omega}_{\text{CC}}^{[\gamma]})$ as a function of $\gamma$ which can be easily computed as in Eq.~(\ref{eq:quantum capacity double maximCE}) that, following the same reasoning that led us to Eq.~(\ref{eq:quantum capacity double maxim323}), rewrites now as 
\begin{eqnarray}\label{eq:quantum capacity double maximCEnew}
Q_E({\Omega}_{\text{CC}}^{[\gamma]})  &=& \max_{ p\in[0,1]}\Big\{ H_2(p) + {(1-p)} \log_2 d_{\text{B}} \nonumber  \\
  && +\frac{1}{2}  \max_{\hat{\tau}_{\text{AA}} \in  \mathfrak{S}_{{\text{A}}} }  \Big\{  p S\left(  
  \hat{\tau}_{\text{AA}}\right)  + p S\left(  
 {\Omega}_{\text{AA}}^{[\gamma]}[\hat{\tau}_{\text{AA}}]\right) \label{QB}  \nonumber  \\ \nonumber 
  && - S(p{\Omega}_{\text{AA}}^{[1-\gamma]}[ \hat{\tau}_{\text{AA}}]+  (1-p)  |0_\text{A}\rangle \! \langle 0_\text{A}| )  \Big\}\Big\}\;. \nonumber \\
\end{eqnarray} 

Again, it is worth mentioning that in the region $1/2\leq \gamma \leq 1$, for $d>2$, the channels are not degradable nor antidegradable but still we are able to compute exactly $Q$ by exploitation of coinciding upper and lower bounds.\\

\subsection{MAD channel plus block dephasing } \label{application1} 
As a final example we now consider the capacity of channels obtained by composing the MAD transformations introduced in the previous section with 
the dephasing channels $\Delta^{(\kappa)}_{\text{CC}}$ that acts over the non diagonal blocks, as shown in Eq.~(\ref{defKappa}), i.e. the maps 
\begin{eqnarray} 
 \Omega_{\text{CC}}^{[\gamma](\kappa)}\equiv  \Delta^{(\kappa)}_{\text{CC}} \circ \Omega_{\text{CC}}^{[\gamma]}=
 \Omega_{\text{CC}}^{[\gamma]}\circ   \Delta^{(\kappa)}_{\text{CC}} \;. 
\end{eqnarray}
As usual let us start with some preliminary observations. 
We can invoke the DPI for the quantum capacity and the internal composition rules of the sets $\Omega_{\text{CC}}^{[\gamma]}$ and 
$\Delta^{(\kappa)}_{\text{CC}}$. With those we can establish the quantum capacities of $\Omega_{\text{CC}}^{[\gamma](\kappa)}$
to be monotonically decreasing in $\gamma$ and monotonically increasing in $|\kappa|$, i.e. 
\begin{eqnarray} \label{fdsd1212f} 
Q( \Omega_{\text{CC}}^{[\gamma](\kappa)})  \geq \max\{  Q( \Omega_{\text{CC}}^{[\gamma'](\kappa)}),
Q( \Omega_{\text{CC}}^{[\gamma](\kappa')})\} 
  \;, 
\end{eqnarray} 
for all $\gamma\leq \gamma'$ and for all $|\kappa|\geq |\kappa'|$. 
Furthermore, again from DPI, it follows that  the quantum capacity of 
$\Omega_{\text{CC}}^{[\gamma](\kappa)}$ is always smaller than or equal to the corresponding value associated with the MAD channel 
$\Omega_{\text{CC}}^{[\gamma]}$, as well as the quantum capacity of $\Delta^{(\kappa)}_{\text{CC}}$ we computed
in Sec.~\ref{Sec:PURELYDEP}, i.e. 
\begin{eqnarray} \label{fdsdf} 
Q( \Omega_{\text{CC}}^{[\gamma](\kappa)})  \leq  \min \{ Q( \Omega_{\text{CC}}^{[\gamma]}) ,Q( \Delta_{\text{CC}}^{(\kappa)}) \} \;. 
\end{eqnarray} 
In particular for $\kappa=0$ (full dephasing), from Eq.~(\ref{DSCAP}) we get 
\begin{eqnarray}
Q( \Omega_{\text{CC}}^{[\gamma](0)}) =  
\max \{ 
Q( \Omega_{\text{AA}}^{[\gamma]}) ,
\log_2 d_{\text{B}} \} \;,
\label{DSCAP112} 
\end{eqnarray}
which, considering that  the capacity $Q( \Omega_{\text{AA}}^{[\gamma]})$ of the  qubit ADC channel 
$\Omega_{\text{AA}}^{[\gamma]}$  is always upper bounded by 1. It is clearly also always smaller than  or equal to the lower bound in Eq.~(\ref{lowerbbb}) of $Q({\Omega}_{\text{CC}}^{[\gamma]})$ as well as smaller than or equal to the value of 
$Q( \Delta_{\text{CC}}^{(0)})$ given in Eq.~(\ref{QK0}).

\begin{figure}[t!]
  \includegraphics[width=\linewidth]{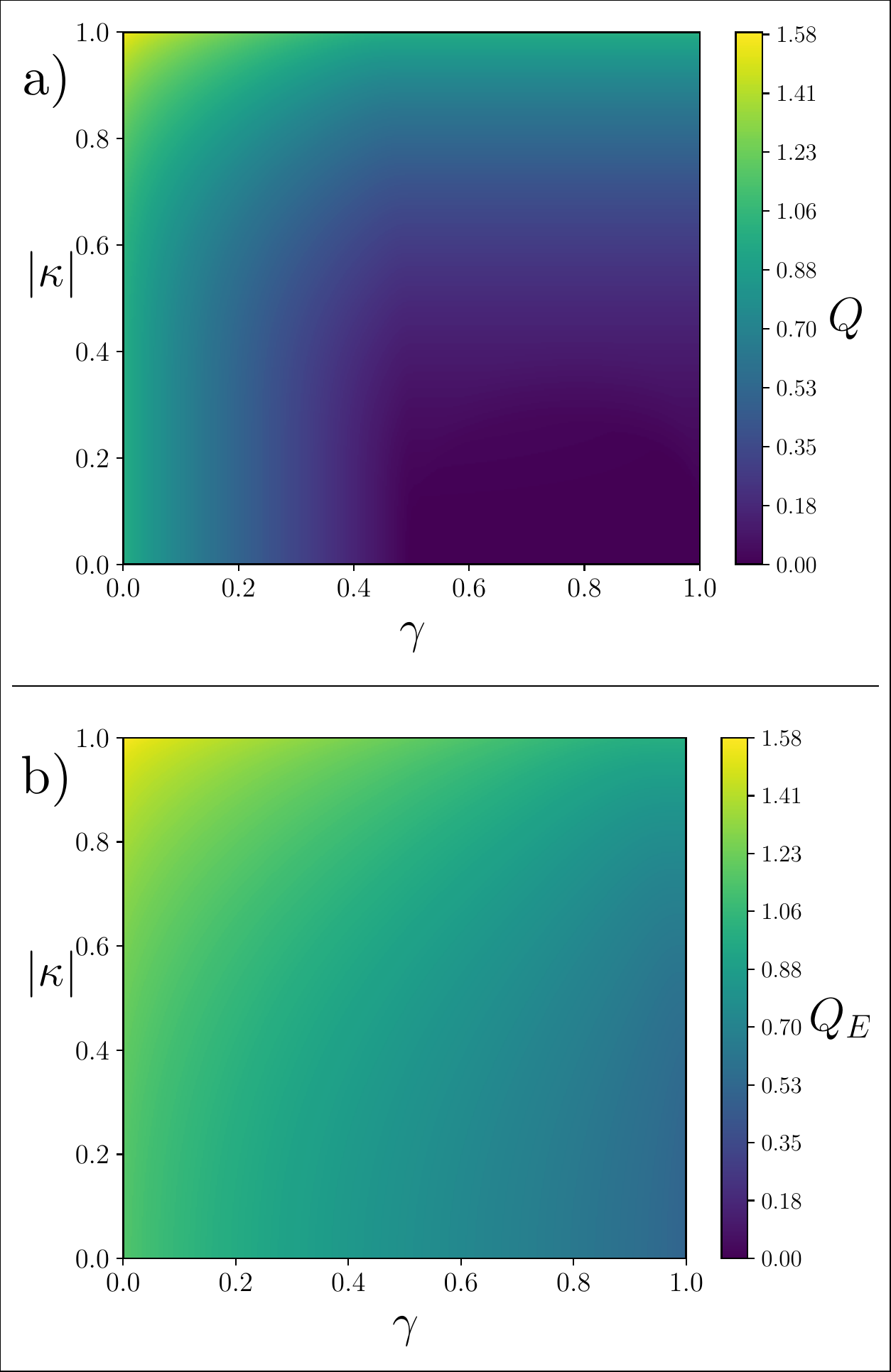}
  \caption{\textbf{a)} Quantum capacity of the channel 
  $\Omega_{\text{CC}}^{[\gamma](\kappa)}$
 w.r.t. the damping parameter $\gamma$ and the dephasing parameter $|\kappa|$. Notice that the map is degradable if and only
 if $\gamma\leq 1/2$. For $\gamma\geq 1/2$ the capacity no longer depends upon $\gamma$ and it is given by Eq.~(\ref{fdsd1212fsadf}). \textbf{b)} Entanglement assisted quantum capacity of the channel 
  $\Omega_{\text{CC}}^{[\gamma](\kappa)}$
 w.r.t. the damping parameter $\gamma$ and the dephasing parameter $|\kappa|$.}
  \label{fig:Qcap_gk}
\end{figure}

To compute the exact value of $Q(\Omega_{\text{CC}}^{[\gamma](\kappa)})$ for $\kappa\neq 0$, observe that as $\Omega_{\text{CC}}^{[\gamma](\kappa)}$ shares the same diagonal block terms of $\Omega_{\text{CC}}^{[\gamma]}$. It will enjoy the same degradability properties of the latter -- see Theorem \ref{thm:optimalDeltaM}. In particular this implies that, irrespectively of the value of 
$\kappa$,  $\Omega_{\text{CC}}^{[\gamma](\kappa)}$ is again degradable if and only if 
 $\gamma\leq 1/2$. Accordingly we can express  $Q(\Omega_{\text{CC}}^{[\gamma](\kappa)})$ using the single letter formula in Eq.~(\ref{eq:quantum capacity double maxim}). In Fig.~\ref{fig:Qcap_gk} a) we report the solution for the case 
 $d_{\text{C}}=3$ obtained by solving numerically the optimization over the input state~$\hat{\tau}_{\text{AA}}$ -- see Appendix~\ref{last:app} for details.  
 
 To obtain the value of $Q(\Omega_{\text{CC}}^{[\gamma](\kappa)})$ also for $1/2\leq \gamma \leq 1$, where the channel
 is explicitly not degradable, we resort to 
produce coinciding upper and lower bounds for such a quantity. Specifically we notice that, irrespectively of the value of $\gamma$,  if we restrict the possible input states to the subspace spanned by $\ket{0}_{\text{C}}$, $\ket{2}_{\text{C}}$ we see that $\Omega_{\text{CC}}^{[\gamma](\kappa)}$ acts just like the qubit dephasing channel. Its quantum capacity corresponds to the value given in Eq.~(\ref{DEFQ}) computed at $d_{\text{A}}=1$~\cite{DEVSHOR} and 
which gives our lower bound, i.e. 
\begin{eqnarray} \label{eq:qubit dephasing}
Q(\Omega_{\text{CC}}^{[\gamma](\kappa)}) \geq 1-H_2((1-|\kappa|)/2)\;.\end{eqnarray} 
An upper bound for $Q(\Omega_{\text{CC}}^{[\gamma](\kappa)})$ for $\gamma>1/2$ instead directly follows from Eq.~(\ref{fdsd1212f}) in the form 
\begin{eqnarray} \label{fdsd1212fa} 
Q( \Omega_{\text{CC}}^{[\gamma](\kappa)})  \leq   Q( \Omega_{\text{CC}}^{[1/2](\kappa)})
  \;. 
\end{eqnarray} 
Now we compute $Q( \Omega_{\text{CC}}^{[\gamma](\kappa)})$ $\forall \, \kappa$ at $\gamma=1/2$ and numerically we verify that it coincides with Eq.~(\ref{eq:qubit dephasing}). Accordingly we can conclude that 
\begin{eqnarray} \label{fdsd1212fsadf} 
Q( \Omega_{\text{CC}}^{[\gamma](\kappa)})  =    1-H_2((1-|\kappa|)/2  
  \;, \qquad \forall \gamma \geq 1/2\;,\nonumber \\
\end{eqnarray} 
as reported in Fig.~\ref{fig:Qcap_gk} a). As before, it is worth mentioning that in the region $1/2\leq \gamma \leq 1$ the channels are not degradable nor antidegradable but still we are able to compute exactly $Q$ by exploitation of coinciding upper and lower bounds.\\

Finally we perform the maximization in Eq.~(\ref{eq:quantum capacity double maximCE}), which gives us $Q_E( \Omega_{\text{CC}}^{[\gamma](\kappa)})$, reported in Fig.~\ref{fig:Qcap_gk} b).

\section{Conclusions}\label{conclusione} 

We firstly introduced the new class of \textit{Partially Coherent Direct Sum } channels. We showed that an explicit and compact formula for the quantum capacity and entanglement assisted quantum capacity is attainable, given suitable degradability conditions of the sub-blocks channels. Since for degradable channels quantum capacity $Q$ and private classical capacity $C_p$ are equivalent \cite{PRIVATE}, the degradability provides us also the latter. Since the expression of $Q_E$ differs from the entanglement assisted classical capacity $C_E$ just by a factor $1/2$ \cite{ENT_ASS1, ENT_ASS2}, given the former we have immediately the latter. We are also able to exhibit upper and lower bounds which, in some occasions, also allow us to state exactly the quantum capacity of non degradable channels. We applied the results to instances of the purely dephasing channels, qubit ADC and combinations of the two. The choice of a qubit ADC made here though was just adopted for the sake of simplicity. The same approach can be straightforwardly applied to higher dimensions ADC when known to be degradable \cite{QUTRIT_ADC} or in general to ``extend'' any other finite dimensional degradable channel. The new approach is immediately generalizable to PCDS composed by $n>2$ block channels. Moreover, since the maximization is reduced to sub-blocks, the overall problem complexity is considerably lower, making a large class of higher dimensional noisy channels capacities accessible.

We acknowledge support from PRIN 2017 ``Taming complexity with quantum strategies".

\appendix

\section{Necessary and sufficient conditions for PCDS quantum channels} \label{APPANEC} 

Here we discuss necessary and sufficient conditions for a quantum channel $\Phi_{\text{CC}} \in {\cal M}^{\text{(cpt)}}_{{\text{C}} \rightarrow {\text{C}}}$ to admit the PCDS block-structure defined in Eq.~(\ref{D2}). 
We start by observing that when introducing this special decomposition 
we did not explicitly require the diagonal blocks  $\Phi_{\text{AA}}$ and  $\Phi_{\text{BB}}$ to be CPT (indeed we merely asked  them  to be elements of the
super-operators sets  ${\cal M}_{{\text{A}} \rightarrow {\text{A}}}$ and ${\cal M}_{{\text{B}} \rightarrow {\text{B}}}$). This
property however is automatically imposed by the CPT constraint on $\Phi_{\text{CC}}$ -- see the derivation that follows.

We now give an explicit proof of Theorem~\ref{thm:necandsuff}. 
First of all, assume that the element ${\hat{M}}_{\text{CC}}^{(j)}$ of the Kraus set 
$\{ {\hat{M}}_{\text{CC}}^{(j)}\}_j$ of $\Phi_{\text{CC}}$ fulfills the identity in Eq.~(\ref{KRAUS}). Accordingly for all ${\hat{\Theta}}_{\text{CC}}\in {\cal{L}}_{{\text{C}} \rightarrow {\text{C}}}$ we can write 
\begin{eqnarray} 
\Phi_{\text{CC}} [{\hat{\Theta}}_{\text{CC}} ]  &=&  \sum_j {\hat{M}}_{\text{CC}}^{(j)} {\hat{\Theta}}_{\text{CC}}
{\hat{M}}_{\text{CC}}^{(j)\dag} \nonumber \\
&=&   \sum_j  \left(\sum_{\text{Y}= \text{A,B}} {\hat{M}}_{\text{YY}}^{(j)}\right) {\hat{\Theta}}_{\text{CC}}  
\left(\sum_{\text{X}= \text{A,B}}
{\hat{M}}_{\text{XX}}^{(j)\dag}\right) \nonumber \\
&=& \sum_j  \sum_{\text{X,Y}= \text{A,B}} {\hat{M}}_{\text{YY}}^{(j)} {\hat{\Theta}}_{\text{YX}}  
{\hat{M}}_{\text{XX}}^{(j)\dag} \;, 
\label{es1} 
\end{eqnarray} 
which can be cast into the form of Eq.~(\ref{BRIEF}) with the super-operators 
$\Phi_{\text{AA}}$, $\Phi_{\text{AA}}$, $\Phi^{\text{(off)}}_{\text{AB}}$, $\Phi^{\text{(off)}}_{\text{BA}}$  defined as in Eqs.~(\ref{OFFDIAG}) and 
(\ref{OFFDIAG1}). 
Notice in particular that with this choice, for ${\text{X}= \text{A,B}}$ 
 the diagonal component
writes 
\begin{eqnarray} 
\Phi_{\text{XX}} [\cdots]  =  \sum_j {\hat{M}}_{\text{XX}}^{(j)}  \cdots {\hat{M}}_{\text{XX}}^{(j)\dag}\;,\label{DIAG} 
\end{eqnarray} 
with the operators  ${\hat{M}}_{\text{XX}}^{(j)}$ fulfilling the constraints
\begin{eqnarray} 
 \sum_j {\hat{M}}_{\text{XX}}^{(j)\dag} {\hat{M}}_{\text{XX}}^{(j)} &=&  
 {\hat{P}}_{\text{XX}} \sum_j {\hat{M}}_{\text{CC}}^{(j)\dag} {\hat{M}}_{\text{CC}}^{(j)}  {\hat{P}}_{\text{XX}}\nonumber\\
&=& {\hat{P}}_{\text{XX}}  \hat{I}_{\text{CC}}
{\hat{P}}_{\text{XX}} =  {\hat{P}}_{\text{XX}}\;.
\end{eqnarray} 
This implies that 
 $\{ {\hat{M}}_{\text{XX}}^{(j)}\}_j$ is a proper Kraus set for a map action on X, i.e. that  $\Phi_{\text{XX}}$ is indeed a CPT element of ${\cal M}_{{\text{X}} \rightarrow {\text{X}}}$, as anticipated in the introduction of the
present section.  

Consider now the reverse property, i.e. assume that  exist 
$\Phi_{\text{AA}}\in {\cal M}_{{\text{A}} \rightarrow {\text{A}}}$, $\Phi_{\text{BB}}\in {\cal M}_{{\text{B}} \rightarrow {\text{B}}}$, $\Phi^{\text{(off)}}_{\text{AB}}\in {\cal M}^{\text{(off)}}_{{\text{B}} \rightarrow {\text{A}}}$, and $\Phi_{\text{BA}}^{\text{(off)}}\in {\cal M}^{\text{(off)}}_{{\text{A}} \rightarrow {\text{B}}}$ 
such that 
Eq.~(\ref{D2}) holds true for all possible choices of $\hat{\Theta}_{\text{CC}}\in {\cal{L}}_{{\text{C}} \rightarrow {\text{C}}}$.
Observe then that this in particular  implies 
  \begin{eqnarray} \label{DFA} 
0&=& {\hat{P}}_{\text{BB}}  \Phi_{\text{CC}}[ {\hat{P}}_{\text{AA}}] {\hat{P}}_{\text{BB}}  = \sum_{j} {\hat{P}}_{\text{BB}}
M_{\text{CC}}^{(j)} {\hat{P}}_{\text{AA}} M_{\text{CC}}^{(j)\dag}{\hat{P}}_{\text{BB}}  \nonumber \\ \nonumber 
&\hspace{-20pt}=&\hspace{-10pt}  \sum_{j}  \left( {\hat{P}}_{\text{BB}}
M_{\text{CC}}^{(j)} {\hat{P}}_{\text{AA}} \right) \left( {\hat{P}}_{\text{BB}}
M_{\text{CC}}^{(j)} {\hat{P}}_{\text{AA}} \right)^{\dag}  = \sum_{j}   |  M_{BA}^{(j)} |^2  \;. 
\end{eqnarray} 
Which  is verified if and only if  
\begin{eqnarray} 
 {\hat{M}}_{\text{BA}}^{(j)} ={\hat{M}}_{\text{AB}}^{(j)} = 0 \qquad \forall j\;,
\end{eqnarray} 
or equivalently if and only if Eq.~(\ref{KRAUS}) holds true. $\square$

\section{Complementary maps via Stinespring dilation} \label{APPA} 

Given $\Phi_{\text{XX}}\in {\cal M}^{\text{(cpt)}}_{{\text{X}} \rightarrow {\text{X}}}$ a generic CPT transformation acting on an arbitrary  system X, we can always express it as 
 \begin{eqnarray}\label{STINESP} 
{\Phi}_{\text{XX}}[\cdots ] = \mbox{Tr}_{\text{E}}[ \hat{U}_{\text{XEXE}} (\cdots \otimes |0_{\text{E}}\rangle \! \langle 0_{\text{E}}| ) \hat{U}^\dag_{\text{XEXE}}]\;,\nonumber \\
\end{eqnarray} 
where E is an auxiliary  (environmental) quantum system,  $\mbox{Tr}_{\text{E}}[ \cdots ]$ is the partial trace over E, $|0_{\text{E}}\rangle$  a pure state vector of the Hilbert space 
${\cal H}_{\text{E}}$ of E, and finally 
$\hat{U}_{\text{XEXE}}$ is
a unitary transformation on ${\cal H}_X\otimes  {\cal H}_{\text{E}}$. For future purposes it  is worth stressing that, by taking the dimensionality of ${\cal H}_{\text E}$ to be sufficiently large, we can make sure that the dependence of the representation in Eq.~(\ref{STINESP}) upon the specific choice of 
${\Phi}_{\text{XX}}$ is completely carried on by just $\hat{U}_{\text{XEXE}}$.  We have then the freedom of fixing
$|0_{\text{E}}\rangle$ irrespectively of the map we want to represent. 
In the above setting a Kraus set for ${\Phi}_{\text{XX}}$ is e.g. obtained in terms of the operators
\begin{eqnarray} \label{KRAUSSETNEW} 
{\hat{M}}_{\text{XX}}^{(j)} =  {\langle} j_{\text{E}}| \hat{U}_{\text{XEXE}}  |0_{\text{E}}\rangle \;,
\end{eqnarray} 
with $\{ |j_{\text{E}}\rangle\}_j$ an orthonormal basis of ${\cal H}_{\text{E}}$.
The
complementary channel of ${\Phi}_{\text{XX}}$ instead can be defined as
the CPT transformation  $\tilde{\Phi}_{\text{EX}}\in {\cal M}^{\text{(cpt)}}_{{\text{X}} \rightarrow {\text{E}}}$ 
that transforms input from X into  output of E via  the mapping 
\begin{equation} \label{COMPPHIstine} 
\tilde{\Phi}_{\text{EX}}[\cdots ] =\hat{V}_{\text{EE}}
 \mbox{Tr}_{\text{X}}[ \hat{U}_{\text{XEXE}} (\cdots \otimes |0_{\text{E}}\rangle \! \langle 0_{\text{E}}| ) \hat{U}^\dag_{\text{XEXE}}]
 \hat{V}^{\dag}_{\text{EE}}\;,
\end{equation} 
where now the partial trace is performed over X. In the above expression  
 $\hat{V}_{\text{EE}}$ is a unitary operator on E that can be chosen freely. We inserted it to explicitly stress that, as already mentioned in the main text, the complementary channel of a CPT map is defined up to a unitary rotation on the environmental system of the model. Anyway, unless explicitly stated, hereafter we shall assume $\hat{V}_{\text{EE}}$ to be the identity operator -- notice that under this assumption, thanks to Eq.~(\ref{KRAUSSETNEW}),
Eq.~(\ref{COMPPHIstine}) reduces exactly to Eq.~(\ref{COMPPHI}) reported in the main text.

An alternative way to derive Eq.~(\ref{ERE12}) can now be obtained by 
first introducing the unitary operators  $\hat{U}_{\text{AEAE}}\in {\cal L}_{{\text{AE}} \rightarrow {\text{AE}}}$ and $\hat{U}_{\text{BEBE}}\in {\cal L}_{{\text{BE}} \rightarrow {\text{BE}}}$  which
provide, respectively, the Stinespring representations of Eq.~(\ref{STINESP})  of  the diagonal components $\Phi_{\text{AA}}$  and 
$\Phi_{\text{BB}}$ of  the PCDS channel $\Phi_{\text{CC}}$. 
Observe now that while  
 the  unitary operator $\hat{U}_{\text{AEAE}}$ ($\hat{U}_{\text{BEBE}}$)  is formally defined 
on ${\cal H}_{\text{A}} \otimes {\cal H}_{\text{E}}$ (${\cal H}_{\text{B}} \otimes {\cal H}_{\text{E}}$) only, we are allowed to 
extend it to the full space ${\cal H}_{\text{C}} \otimes {\cal H}_{\text{E}}$ by imposing the condition 
$\hat{P}_{{\text{B}}}  \hat{U}_{\text{AEAE}} =  \hat{U}_{\text{AEAE}}\hat{P}_{{\text{B}}}  =0$ (resp.
$\hat{P}_{{\text{A}}}  \hat{U}_{\text{BEBE}} =  \hat{U}_{\text{BEBE}}\hat{P}_{{\text{A}}}  =0$).
With this choice hence we can write the normalization condition for $\hat{U}_{\text{AEAE}}$ and 
$\hat{U}_{\text{BEBE}}$ as 
\begin{eqnarray} 
\hat{U}^\dag_{\text{AEAE}} \hat{U}_{\text{AEAE}} &=& \hat{P}_{\text{AA}} \otimes \hat{I}_{\text{EE}}\;, 
\nonumber \\ \label{QUETA} 
\hat{U}^\dag_{\text{BEBE}} \hat{U}_{\text{BEBE}} &=& \hat{P}_{\text{BB}} \otimes \hat{I}_{\text{EE}}\;, 
\end{eqnarray}
 with the projectors  $\hat{P}_{\text{AA}}$ and $\hat{P}_{\text{AA}}$ playing the role of the identity transformations on  ${\cal H}_{\text{A}}$ and ${\cal H}_{\text{B}}$ respectively. 
 In view of these observations a  Stinespring representation as in Eq.~(\ref{STINESP}) 
for the PCDS channel $\Phi_{\text{CC}}$ 
can now be assigned by adopting the following coupling 
\begin{eqnarray} 
\hat{U}_{\text{CECE}} = \hat{U}_{\text{AEAE}}  +  \hat{U}_{\text{BEBE}}  \;,
\end{eqnarray} 
which is a unitary transformation on ${\cal H}_{\text{C}} \otimes  {\cal H}_{\text{E}}$ thanks to Eq.~(\ref{QUETA}) and Eq.~(\ref{ORTHO}).
Furthermore,  thanks to Eq.~(\ref{KRAUSSETNEW}) automatically fulfills the necessary and sufficient PCDS condition in Eq.~(\ref{KRAUS}).
To verify Eq.~(\ref{ERE12}) now observe that for an arbitrary vector 
  $|\Psi_{\text{C}}\rangle\in{\cal H}_{\text{C}}$ we can write 
\begin{eqnarray} 
\hat{U}_{\text{CECE}}( |\Psi_{\text{C}}\rangle \otimes |0_{\text{E}}\rangle) &=& 
\hat{U}_{\text{AEAE}} ( |\Psi_{\text{A}}\rangle  \otimes |0_{\text{E}}\rangle  ) \nonumber \\ 
& &+\; \hat{U}_{\text{BEBE}}  (|\Psi_{\text{B}}\rangle \otimes |0_{\text{E}}\rangle)\;, \nonumber \\
\end{eqnarray}
where for $\text{X}=\text{A,B}$,  $|\Psi_{\text{X}}\rangle \equiv \hat{P}_{\text{XX}} |\Psi_{\text{C}}\rangle$.
Tracing over C from the above expression we get that the action 
of $\tilde{\Phi}_{\text{EC}}$ on $|\Psi_{\text{C}}\rangle$ can be expressed as 
\begin{equation} 
\tilde{\Phi}_{\text{EC}}[|\Psi_{\text{C}}\rangle \! \langle \Psi_{\text{C}}|] = \tilde{\Phi}_{\text{EA}}[|\Psi_{\text{A}}\rangle \! \langle \Psi_{\text{A}}|] + 
\tilde{\Phi}_{\text{EB}}[|\Psi_{\text{B}}\rangle \! \langle \Psi_{\text{B}}|] \;, \label{ERE1} 
\end{equation}
where for $\text{X}=\text{A,B}$, $\tilde{\Phi}_{\text{EX}}$ is the complementary map of $\Phi_{\text{XX}}$, and 
where we used the fact that $\hat{U}_{\text{AEAE}} ( |\Psi_{\text{A}}\rangle  \otimes |0_{\text{E}}\rangle  )$ lives on ${\cal H}_{\text{A}} \otimes {\cal H}_{\text{E}}$ and therefore has
zero overlap with the  C  components of $\hat{U}_{\text{BEBE}} ( |\Psi_{\text{B}}\rangle  \otimes |0_{\text{E}}\rangle  )$, which instead is on ${\cal H}_{\text{B}} \otimes {\cal H}_{\text{E}}$.

\subsection{Structure of the connecting channels} \label{APPCPTCOND} 
Here we show that if $\Lambda_{\text{EA}}$ and $\Lambda_{\text{EB}}$ entering in 
Eq.~(\ref{DEFCON}) are both CPT then also $\Lambda_{\text{EC}}$ is CPT.
To see this  remember that given $\Phi_{\text{YX}}\in  {\cal M}_{{\text{X}} \rightarrow {\text{Y}}}$ a  super-operator
mapping the system X into Y, it is CPT if and only if it admits 
a Kraus set  formed by operators $\hat{M}_{\text{YX}}^{(j)}$ that fulfill the normalization
condition 
\begin{eqnarray}\label{eq: Kraus deg chan}
\sum_j \hat{M}_{\text{YX}}^{(j)\dag}\hat{M}_{\text{YX}}^{(j)} = \hat{I}_{\text{XX}} \;,
\end{eqnarray} 
with $\hat{I}_{\text{XX}}$ the identity on ${\cal H}_{\text{X}}$, or alternatively the associated projector in case 
${\cal H}_{\text{X}}$ is a sub-space on a larger space.
Consequently, since we assumed by hypothesis that this is the case for  $\Lambda_{\text{EA}}$ and $\Lambda_{\text{EB}}$
appearing in Eq.~(\ref{DEFCON}), it follows that a Kraus set for $\Lambda_{\text{EC}}$ is given by the set $\{\hat{M}_{\text{EA}}^{(j_1)},
\hat{M}_{\text{EB}}^{(j_1)}\}_{j_1,j_2}$. As indeed we have  
\begin{eqnarray} \label{eq: norm Kraus deg chan}
\sum_{j_1} \hat{M}_{\text{EA}}^{(j_1)\dag}\hat{M}_{\text{EA}}^{(j_1)} +\sum_{j_2} \hat{M}_{\text{EB}}^{(j_2)\dag}\hat{M}_{\text{EB}}^{(j_2)} &=& \hat{P}_{\text{AA}} +\hat{P}_{\text{BB}} \nonumber \\
& =&\hat{I}_{\text{CC}}  \;.
\end{eqnarray} 
\subsection{Pure fixed point channels} \label{appPURE} 
Here we show that if the quantum channel ${\Phi}_{\text{AA}}\in
{\cal M}^{\text{(cpt)}}_{{\text{A}} \rightarrow {\text{A}}}$ admits as fixed point 
a pure state $|\Psi^*_{\text{A}}\rangle \in{\cal H}_{\text{A}}$, then condition in Eq.~(\ref{FFd1}) can be
fulfilled by setting $\hat{\rho}^*_{\text{AA}}= |\Psi^*_{\text{A}}\rangle \! \langle \Psi^*_{\text{A}}|$. 
To show this, let us consider the unitary operator 
$\hat{U}_{\text{AEAE}}$ that allows us to express ${\Phi}_{\text{AA}}$ and its associated
complementary channel  ${\Phi}_{\text{EA}}$ in the Stinespring representation given by
Eqs.~(\ref{STINESP}) and 
(\ref{COMPPHIstine}).
The fixed point condition of $|\Psi^*_{\text{A}}\rangle$ imposes us to have 
\begin{eqnarray} 
{\Phi}_{\text{AA}}[ |\Psi^*_{\text{A}}\rangle \!\langle \Psi^*_{\text{A}}|] = |\Psi^*_{\text{A}}\rangle\!\langle \Psi^*_{\text{A}}|\;,
\end{eqnarray}
which can be satisfied if and only if the following identity holds true:
\begin{eqnarray} 
\hat{U}_{\text{AEAE}} |\Psi^*_{\text{A}}\rangle \otimes |0_{\text{E}}\rangle = |\Psi^*_{\text{A}}\rangle \otimes |0'_{\text{E}}\rangle\;,
\end{eqnarray}
with $|0'_E\rangle$ 
being some pure state of E. Accordingly from Eq.~(\ref{COMPPHIstine}) it follows that 
\begin{eqnarray} 
\tilde{\Phi}_{\text{EA}}[ |\Psi^*_A\rangle \! \langle \Psi^*_A|]  = 
{\hat V}_{\text{EE}}  |0'_{\text{E}}\rangle \! \langle 0'_{\text{E}}|{\hat V}^\dag_{\text{EE}} \;,
\end{eqnarray}
 where now we make explicit use of the freedom of redefining  $\tilde{\Phi}_{\text{EA}}$
 up to an arbitrary unitary transformation ${\hat V}_{\text{EE}}$. 
 The condition in Eq.~(\ref{FFd1}) can finally be enforced by simply selecting 
  ${\hat V}_{\text{EE}}$ so that 
  \begin{eqnarray}
 {\hat V}_{\text{EE}}  |0'_{\text{E}}\rangle = |0_{\text{E}}\rangle \;.
 \end{eqnarray} 

\section{Generalization to the multi-block decomposition}\label{GENAPP} 
Consider the case in which 
the Hilbert space of C decomposes in a direct sum of $n$ different subspaces
\begin{eqnarray}
{\cal H}_{\text{C}} = {\cal H}_{\text{A}_1} \oplus {\cal H}_{\text{A}_2} \oplus \cdots \oplus {\cal H}_{\text{A}_n} \;,
\end{eqnarray} 
where for $\ell=1,\cdots,n$, ${\cal H}_{\text{A}_\ell}$ represents a Hilbert space of dimension $d_{\text{A}_\ell}$, with 
\begin{eqnarray} 
d_{\text{C}} = \sum_{\ell=1}^n d_{\text{A}_\ell}\;. 
\end{eqnarray} 
A PCDS CPT channel $\Phi_{\text{CC}}\in {\cal M}^{\text{(cpt)}}_{ \text{C} \rightarrow \text{C}}$ is now defined by the following structural constraint which generalizes the one we presented in Eq.~(\ref{BRIEF}): 
\begin{eqnarray} \Phi_{\text{CC}} = \sum_{\ell=1}^n \Phi_{\text{A}_\ell \text{A}_{\ell}}+ 
\sum_{\ell\neq\ell'=1}^n \Phi^{\text{(off)}}_{\text{A}_\ell \text{A}_{\ell'}}\;, \label{BRIEFn} 
\end{eqnarray} 
with $\Phi_{\text{A}_\ell \text{A}_{\ell}} \in {\cal M}_{ \text{A}_{\ell} \rightarrow \text{A}_{\ell}}$ 
and $\Phi^{\text{(off)}}_{\text{A}_\ell \text{A}_{\ell'}} \in {\cal M}^{\text{(off)}}_{ \text{A}_{\ell} \rightarrow \text{A}_{\ell}}$. 
Following the same derivation we presented for the $n=2$ one can verify that the CPT constraint on 
$\Phi_{\text{CC}}$ imposes all the diagonal terms $\Phi_{\text{A}_\ell \text{A}_{\ell}}$ to be CPT as well. Furthermore
Theorem~\ref{thm:necandsuff} still holds true in the following form
 \begin{thm}
A quantum channel $\Phi_{\text{CC}}$  described by a Kraus set $\{ {\hat{M}}_{\text{CC}}^{(j)}\}_j$
admits the PCDS structure as in Eq.~(\ref{BRIEF}) if and only if  
 \begin{eqnarray} \label{KRAUSn} 
 {\hat{M}}_{\text{CC}}^{(j)} = \bigoplus_{\ell=1}^n {\hat{M}}_{\text{A}_{\ell}\text{A}_{\ell}}^{(j)} \;,  
 \end{eqnarray}
 or equivalently  that ${\hat{M}}_{\text{A}_{\ell}\text{A}_{\ell'}}^{(j)} =0$, for all $j$ and for all $\ell\neq \ell'$.
		\label{thm:necandsuffn}
\end{thm}
In the above expression for all ${\hat{\Theta}}_{\text{CC}}  \in {\cal{L}}_{{\text{C}} \rightarrow {\text{C}}}$ 
we defined
\begin{eqnarray} 
{\hat{\Theta}}_{\text{A}_{\ell}\text{A}_{\ell'}}\equiv \hat{P}_{\text{A}_{\ell}\text{A}_{\ell}} {\hat{\Theta}}_{\text{CC}} \hat{P}_{\text{A}_{\ell'}\text{A}_{\ell'}}\;,
\end{eqnarray} 
with $ \hat{P}_{\text{A}_{\ell}\text{A}_{\ell}} $ being the orthogonal projector on ${\cal H}_{\text{A}_\ell}$.
Accordingly Eqs.~(\ref{OFFDIAG}) and  (\ref{OFFDIAG1})  get replaced by 
\begin{eqnarray} 
\Phi_{\text{A}_\ell \text{A}_{\ell}} [\cdots]  &=&  \sum_j {\hat{M}}_{{\text{A}_{\ell}\text{A}_{\ell}}}^{(j)}  \cdots {\hat{M}}_{{\text{A}_{\ell}\text{A}_{\ell}}}^{(j)\dag}\;,\label{OFFDIAGn} \nonumber \\
\Phi^{\text{(off)}}_{\text{A}_\ell \text{A}_{\ell'}}  [\cdots]  &=&  \sum_j {\hat{M}}_{{\text{A}_{\ell}\text{A}_{\ell}}}^{(j)}  \cdots {\hat{M}}_{{\text{A}_{\ell'}\text{A}_{\ell'}}}^{(j)\dag}\;.\label{OFFDIAG1n} 
\end{eqnarray} 
Similarly Eq.~(\ref{ERE12}) becomes now
\begin{eqnarray}\tilde{\Phi}_{\text{EC}} = \sum_{\ell=1}^n \tilde{\Phi}_{\text{EA}_\ell} \;, 
\end{eqnarray} 
with $\tilde{\Phi}_{\text{EA}_\ell}$ being the complementary channel of ${\Phi}_{\text{A}_\ell,\text{A}_\ell}$. 
Theorem~\ref{thm:optimalDeltaM} instead is replaced by the more general statement 
\begin{thm}
A PCDS  quantum channel $\Phi_{\text{CC}}$ as in Eq.~(\ref{BRIEFn})   is degradable if and only if all its diagonal block terms  
$\Phi_{\text{A}_\ell \text{A}_{\ell}}$   are degradable too.
		\label{thm:optimalDeltaMn}
\end{thm}
It then follows that for $\Phi_{\text{CC}}$ degradable we can express the quantum capacity as 
\begin{eqnarray}
Q(\Phi_{\text{CC}})&=& \max_{P} \max_{\hat{\tau}_{\text{A}_\ell \text{A}_\ell}} \Big\{  H(P) + \sum_{\ell=1}^n 
p_{\ell}  S\left(  
  \Phi_{\text{A}_\ell \text{A}_\ell}[\hat{\tau}_{\text{A}_\ell \text{A}_\ell}]\right)  \nonumber \\
&& - S\left(\sum_{\ell=1}^n 
p_{\ell}  \tilde{\Phi}_{\text{EA}_\ell}[\hat{\tau}_{\text{A}_\ell \text{A}_\ell}]\right)  \Big\}  \;, 
  \label{QBdsd} 
\end{eqnarray} 
with $P$ a generic probability set $\{ p\}_\ell$, 
 $H(P)= -\sum_\ell p_\ell \log p_\ell$ its Shannon entropy, and $\hat{\tau}_{\text{A}_\ell \text{A}_\ell}$ density matrices of ${\cal H}_{\text{A}_\ell}$. 

\section{The channel  $\Omega_{\text{CC}}^{[\gamma](\kappa)}$  for $d_{\text{C}}=3$} \label{last:app}

 When $d_{\text{C}}=3$ a Kraus set of $\Omega_{\text{CC}}^{[\gamma](\kappa)}$
 expressed w.r.t. the canonical base elements $\{ \ket{0_{\text{C}}},\ket{1_{\text{C}}},\ket{2_{\text{C}}}\}$,  
 can be written
as \begin{equation}\label{eq:Kraus13}
\begin{split}
\hat{M}^{(0)}_{\text{CC}}=&\begin{pmatrix}
1 & 0 & 0\\
0 & \sqrt{1-\gamma} & 0\\
0 & 0 & {\kappa}^*
\end{pmatrix}\;, \quad
\hat{M}^{(1)}_{\text{CC}}=\begin{pmatrix}
0 & \sqrt{\gamma} & 0\\
0 & 0 & 0\\
0 & 0 & 0
\end{pmatrix}\;, \\
&\qquad\quad\hat{M}^{(2)}_{\text{CC}}=\begin{pmatrix}
0 & 0 & 0\\
0 & 0 & 0\\
0 & 0 & \sqrt{1-|\kappa|^2}
\end{pmatrix}.
\end{split}
\end{equation}
This leads to 
\begin{equation}
\scalebox{0.95}{$
\Omega_{\text{CC}}^{[\gamma](\kappa)}[\hat{\rho}_{\text{CC}}]=\begin{pmatrix}
\rho_{00}+\gamma \rho_{11}& \sqrt{1-\gamma}\rho_{01} & \kappa\rho_{02}\\
\sqrt{1-\gamma}\rho_{01}^* & (1-\gamma)\rho_{11} & \kappa\sqrt{1-\gamma}\rho_{12}\\
\kappa^*\rho_{02}^* & \kappa^*\sqrt{1-\gamma}\rho_{12}^* & \rho_{22}
\end{pmatrix}$},
\end{equation}
where for $i,j=0,1,2$ we set $\rho_{ij}=\langle i_{\text{C}} |\hat{\rho}_{\text{CC}}| j_{\text{C}}\rangle$, and 
\begin{equation}\label{eq:comp chan amp dep}
\scalebox{0.95}{$
\hspace{-15pt}\tilde{\Omega}_{\text{EC}}^{[\gamma](\kappa)}[\hat{\rho}_{\text{CC}}]
=\begin{pmatrix}
1-\gamma \rho_{11}+|\kappa |^2\rho_{22}& \sqrt{\gamma}\rho_{01} & \kappa^*\sqrt{(1-|\kappa |^2)}\rho_{22}\\
\sqrt{\gamma}\rho_{01}^* & \gamma\rho_{11} & 0\\
\kappa\sqrt{(1-|\kappa |^2)}\rho_{22}^* & 0 & (1-|\kappa |^2)\rho_{22}
\end{pmatrix}$},
\end{equation}
for the complementary map defined on a Hilbert space spanned by the vectors $\{ \ket{0_{\text{E}}},\ket{1_{\text{E}}},\ket{2_{\text{E}}}\}$. 

Notice that we can express the input states $\hat{\rho}_{\text{CC}}$ in terms of the ${\hat{\tau}}_{\text{AA}}$ and ${\hat{\tau}}_{\text{BB}}$ density matrices
as in Eq.~(\ref{ineq1}).  Eq.~(\ref{eq:comp chan amp dep}) then can be equivalently written as 
\begin{eqnarray}\label{ECCO2asdf} 
\tilde{\Omega}_{\text{EC}}^{[\gamma](\kappa)}[\hat{\rho}_{\text{CC}}] = 
p \tilde{\Omega}_{\text{EA}}^{[\gamma]}[{\hat{\tau}}_{\text{AA}} ] + (1-p) |v^{(\kappa)}_{\text{E}}\rangle \!  \langle v^{(\kappa)}_{\text{E}}|\;, \nonumber \\  
\end{eqnarray} 
with $\tilde{\Omega}_{\text{EA}}^{[\gamma]}$, the complementary channel of the MAD channel
${\Omega}_{\text{AA}}^{[\gamma]}$.
$\tilde{\Omega}_{\text{EA}}^{[\gamma]}$ is defined by the $2\times2$ matrix
\begin{equation}
\scalebox{1}{$
\tilde{\Omega}_{\text{EA}}^{[\gamma]}[{\hat{\tau}}_{\text{AA}} ]
=\begin{pmatrix}
1-\gamma_1 \tau_{11}& \sqrt{\gamma_1}\tau_{01} \\
\sqrt{\gamma_1}\tau_{01}^* & \gamma_1\tau_{11} 
\end{pmatrix}$},
\end{equation}
 on the Hilbert space spanned by the vectors 
$|0_{\text{E}}\rangle$ and $|1_{\text{E}}\rangle$. 
 $|v^{(\kappa)}_{\text{E}}\rangle$ in Eq.~\ref{ECCO2asdf} is defined as 
\begin{eqnarray}
|v^{(\kappa)}_{\text{E}}\rangle \equiv \kappa |0_{\text{E}}\rangle + \sqrt{1 - |\kappa|^2} |2_{\text{E}}\rangle \label{defv1} \;,
\end{eqnarray} 
which has the same structure of the state in Eq.~(\ref{defv}) but involves different basis vectors in order to account for the presence of the 
MAD contribution to ${\Omega}_{\text{CC}}^{[\gamma](\kappa)}$.

Now, considering the fact that both ADC and dephasing are covariant w.r.t. the action of the group of diagonal orthogonal matrices \cite{QUTRIT_ADC, DEPH_MEMO}, the maximization of the coherent information is attained by exploring only diagonal states. Consequently, from Eq.~(\ref{eq:quantum capacity double maxim}) and Eq.~(\ref{ECCO2asdf}) the quantum capacity is obtained by:

\begin{equation}
\begin{split}
Q(\Omega_{\text{CC}}^{[\gamma](\kappa)})&=\max\limits_{p\in[0,1]} \big\{ H_2(p)+
 \max\limits_{\tau_{11}\in[0,1]}\{ pH_2(\gamma\tau_{11})\\
& + l_0\log_2l_0+l_+\log_2l_++l_-\log_2l_-\}\big\}
\end{split}
\end{equation}
where 
\begin{equation}
\begin{cases}
l_0=p\tau_{11}\gamma\\
l_+=\frac{1}{2}( 1 -p\gamma \tau_{11} +\\
\quad\qquad\sqrt{ 4(1-p)p(|\kappa|^2-1)(1-\gamma \tau_{11}) + (1-p\gamma\tau_{11})^2 })\\
l_-=\frac{1}{2}( 1 -p\gamma \tau_{11} -\\
\quad\qquad\sqrt{ 4(1-p)p(|\kappa|^2-1)(1-\gamma \tau_{11}) + (1-p\gamma\tau_{11})^2 }).
\end{cases}
\end{equation}
Notice how this method allows us to reduce to just 2 the number of parameters involved in the maximization, compared to the at least 8 needed for a generic qutrit state.

\clearpage

\bibliographystyle{apsrev}
\bibliography{directsum_QUANTUM}

\end{document}